\documentclass[sigconf, nonacm]{acmart}

\newcommand\vldbdoi{10.14778/3485450.3485460}
\newcommand\vldbpages{98 - 111}
\newcommand\vldbvolume{15}
\newcommand\vldbissue{1}
\newcommand\vldbyear{2021}
\newcommand\vldbauthors{\authors}
\newcommand\vldbtitle{DeepEverest: Accelerating Declarative Top-K Queries for Deep Neural Network Interpretation} 
\newcommand\vldbavailabilityurl{https://github.com/uwdb/deepeverest}
\newcommand\vldbpagestyle{plain}

\usepackage[noend]{algpseudocode}
\usepackage{algorithm}
\usepackage{bbm}
\usepackage{graphicx}
\usepackage[compress]{cleveref}
\usepackage{mathrsfs}
\usepackage{bm}
\usepackage{balance}
\usepackage{outlines}
\usepackage{svg}
\usepackage{grffile}
\usepackage[caption=false]{subfig}
\usepackage{xspace}
\usepackage{enumitem}
\usepackage{xcolor}
\usepackage{array}
\usepackage{nccmath}

\usepackage{xpatch}
\xpatchcmd{\NCC@ignorepar}{%
\abovedisplayskip\abovedisplayshortskip}
{%
\abovedisplayskip\abovedisplayshortskip%
\belowdisplayskip\belowdisplayshortskip}
{}{}

\graphicspath{./figures/}

\crefname{section}{Section}{Sections}
\Crefname{section}{Section}{Sections}
\crefname{figure}{Figure}{Figures}
\Crefname{figure}{Figure}{Figures}
\crefname{subfigure}{Figure}{Figures}
\Crefname{subfigure}{Figure}{Figures}
\crefname{algorithm}{Algorithm}{Algorithms}
\Crefname{algorithm}{Algorithm}{Algorithms}
\crefname{table}{Table}{Tables}
\Crefname{table}{Table}{Tables}
\crefrangelabelformat{subfigure}{#3#1#4--#5(\crefstripprefix{#1}{#2}#6}

\newcommand{\revisioncolor}{black}
\newcommand{\revision}[1]{{\color{\revisioncolor} #1}}

\newcommand{\minorcolor}{black}
\newcommand{\minor}[1]{{\color{\minorcolor} #1}}

\newcommand{\cifar}{\textit{CIFAR10-VGG16}\xspace}
\newcommand{\imagenet}{\textit{ImageNet-ResNet50}\xspace}

\makeatletter
\newcommand{\my@arrow}[1]{\ooalign{$#1-\mkern-5mu-$\cr\hidewidth$#1>$}}
\newcommand{\myarrow}{\mathrel{\mathpalette\my@arrow\relax}}
\makeatother

\begin{document}

\title{DeepEverest: Accelerating Declarative Top-K Queries for\\Deep Neural Network Interpretation}
\subtitle{Technical Report}

\author{Dong He, Maureen Daum, Walter Cai, Magdalena Balazinska}
\email{{donghe, mdaum, walter, magda}@cs.washington.edu}
\affiliation{%
  \institution{Paul G. Allen School of Computer Science \& Engineering, University of Washington}
}

\begin{sloppypar}
\begin{abstract}
We \minor{design}, implement, and evaluate DeepEverest, a system for the efficient execution of \textit{interpretation by example} queries over the activation values of a deep neural network. 
DeepEverest consists of an efficient indexing technique and a query execution algorithm with various optimizations. 
We prove that the proposed query execution algorithm is instance optimal. 
Experiments with our prototype show that DeepEverest, using less than $20\%$ of the storage of full materialization, significantly accelerates individual queries by up to \revision{$63{\times}$} and consistently outperforms other methods on multi-query workloads that simulate DNN interpretation processes. 
\end{abstract}
\end{sloppypar}

\maketitle

\pagestyle{\vldbpagestyle}
\begingroup\small\noindent\raggedright\textbf{Reference Format}\\
This is an extended technical report of the following paper:\\
\vldbauthors. \vldbtitle. PVLDB, \vldbvolume(\vldbissue): \vldbpages, \vldbyear.\\
\href{https://doi.org/\vldbdoi}{doi:\vldbdoi}
\endgroup
\begingroup
\renewcommand\thefootnote{}\footnote{\noindent
 This work is licensed under the Creative Commons BY-NC-ND 4.0 International License. Visit \url{https://creativecommons.org/licenses/by-nc-nd/4.0/} to view a copy of this license. Copyright is held by the owner/author(s).
}\addtocounter{footnote}{-1}\endgroup

\ifdefempty{\vldbavailabilityurl}{}{
\vspace{0.5em}
\begingroup\small\noindent\raggedright\textbf{Artifact Availability}\\
The source code, data, and/or other artifacts have been made available at \url{\vldbavailabilityurl}.
\endgroup
}

\begin{sloppypar}

\newcommand{\exampleNeuralPartitionIndex}{
    \begin{figure}[t!]
        \captionsetup{font={color=\revisioncolor}}
        \centering
        \includegraphics[width=\linewidth]{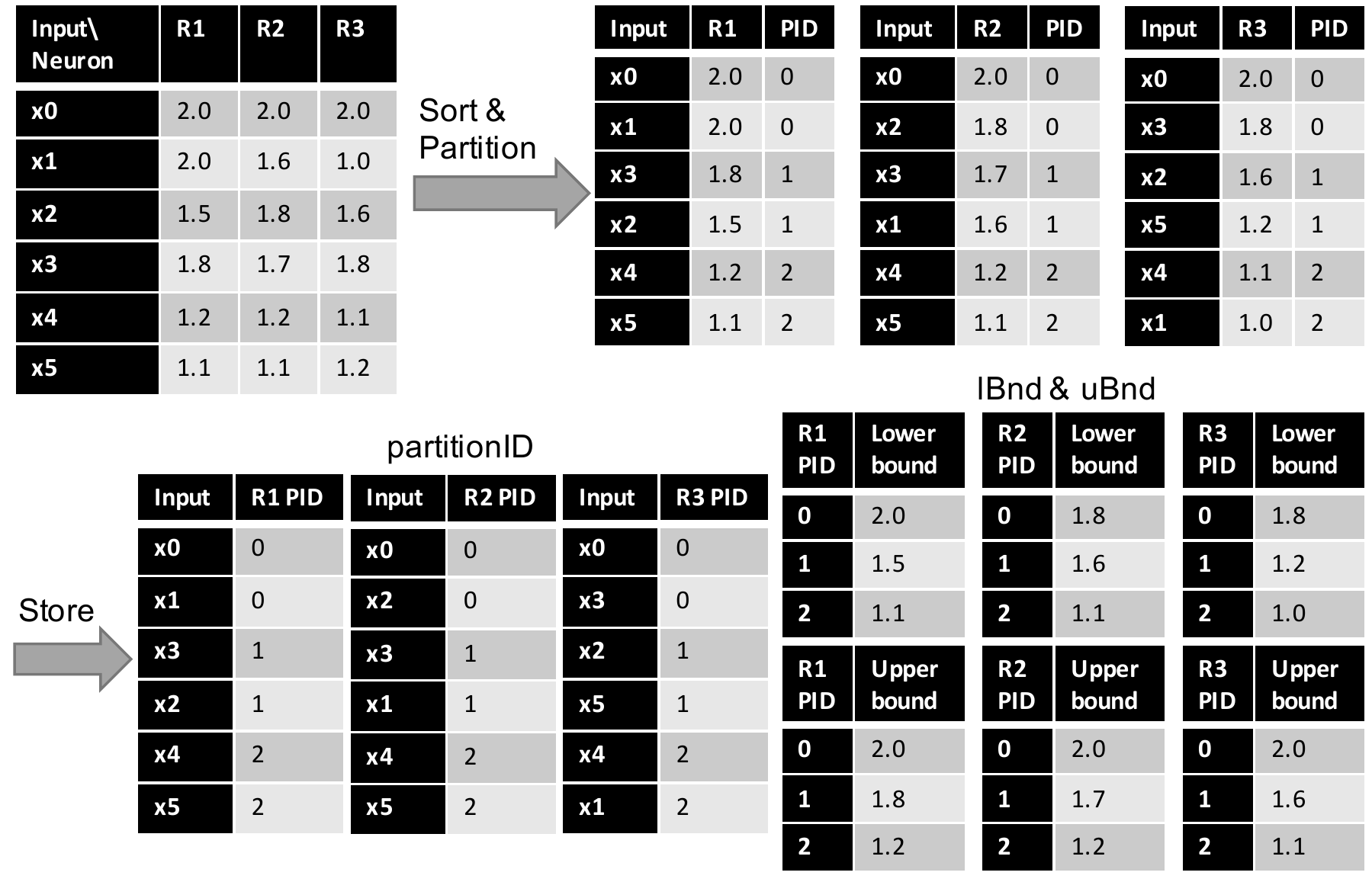}
        \vspace{-2em}
        \caption{An example of building the Neural Partition Index of three neurons, $R1, R2, R3$, for six inputs, $x0, \dots, x5$.}
        \vspace{-1.1em}
        \label{fig:exampleNeuralPartitionIndex}
        \vspace{-0.2em}
    \end{figure}
}

\newcommand{\exampleDParOrd}{
    \begin{figure}[t!]
        \captionsetup{font={color=\revisioncolor}}
        \centering
        \includegraphics[width=0.88\linewidth]{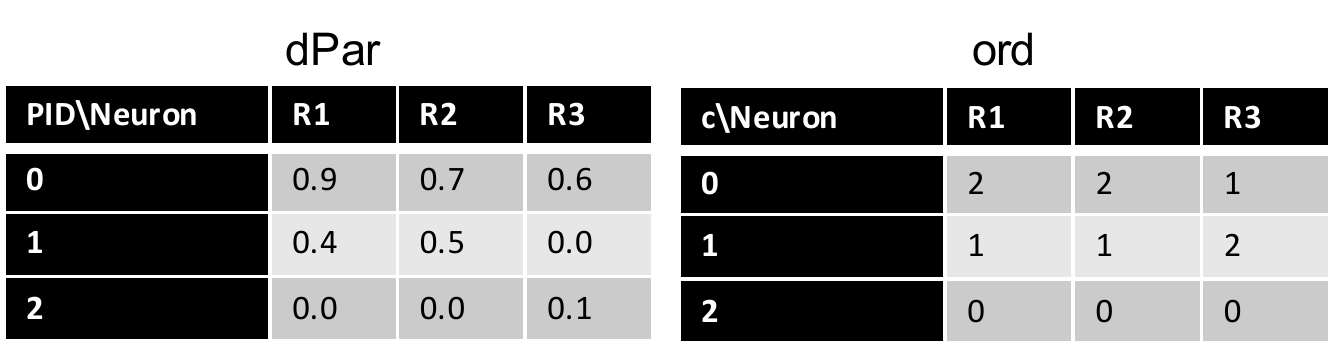}
        \vspace{-1.1em}
        \caption{$dPar$ and $ord$ for the execution of NTA for the example query that finds the most similar inputs to $x5$ based on the activations of $\{R1, R2, R3\}$. For neuron $i$, $c$ is the index of the \texttt{PID}s in $ord(i)$, e.g., when $c {=} 0$, $ord(R1, c) {=} 2$.}
        \vspace{-2.5em}
        \label{fig:exampleDParOrd}
    \end{figure}
}

\newcommand{\exampleToRunDistTop}{
    \begin{figure}[t!]
        \captionsetup{font={color=\revisioncolor}}
        \centering
        \subfloat[Build $toRun$]{\includegraphics[width=\linewidth]{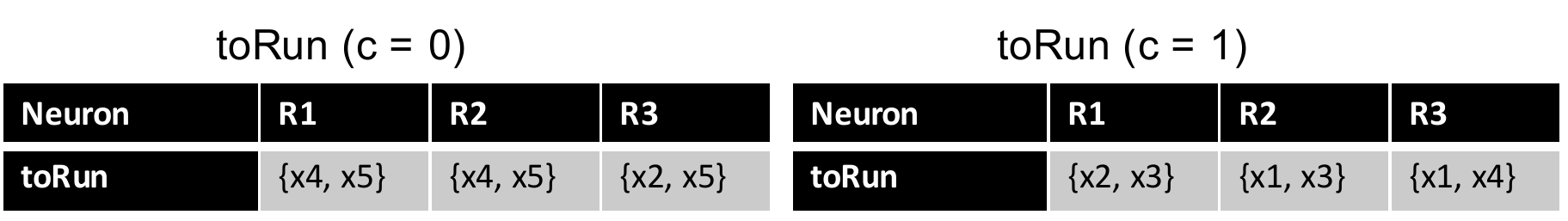}%
        \label{subfig:toRun}}%
        \vspace{-1.4em}

        \subfloat[\revision{Perform DNN inference to compute $dist$ and update $top$}]{\includegraphics[width=\linewidth]{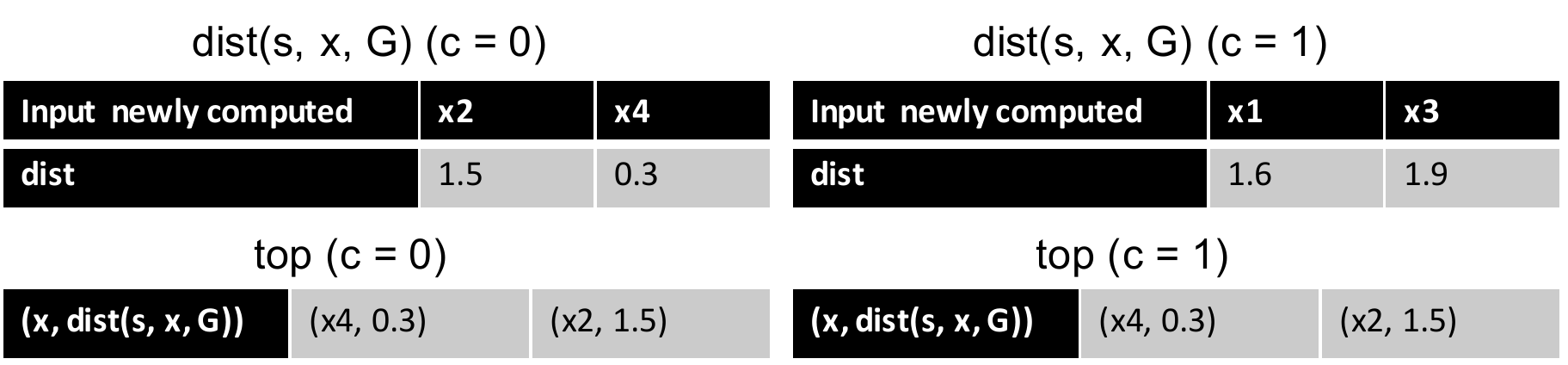}%
        \label{subfig:dist}}%
        \vspace{-1.0em}

        \subfloat[\revision{Check termination}]{\includegraphics[width=\linewidth]{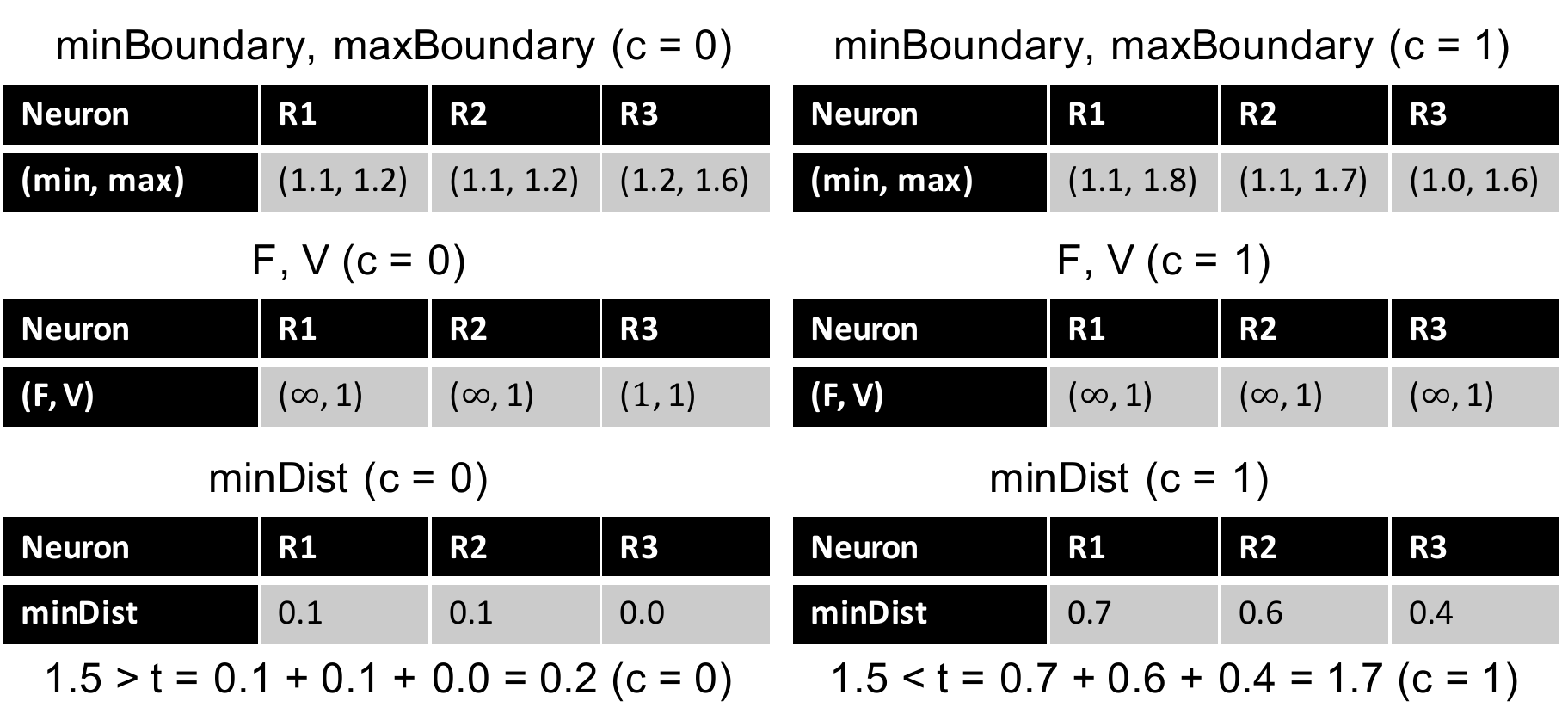}%
        \label{subfig:termination}}%
        \vspace{-1.1em}
        \caption{Intermediate variables for the execution of NTA for the example query. The values when $c{=}0$ are shown on the left, and the values when $c{=}1$ are shown on the right.}
        \vspace{-0.8em}
        \label{fig:exampleToRunDistTop}
    \end{figure}
}

\newcommand{\exampleMaximumActivationIndex}{
    \begin{figure}[t!]
        \captionsetup{font={color=\revisioncolor}}
        \centering
        \includegraphics[width=\linewidth]{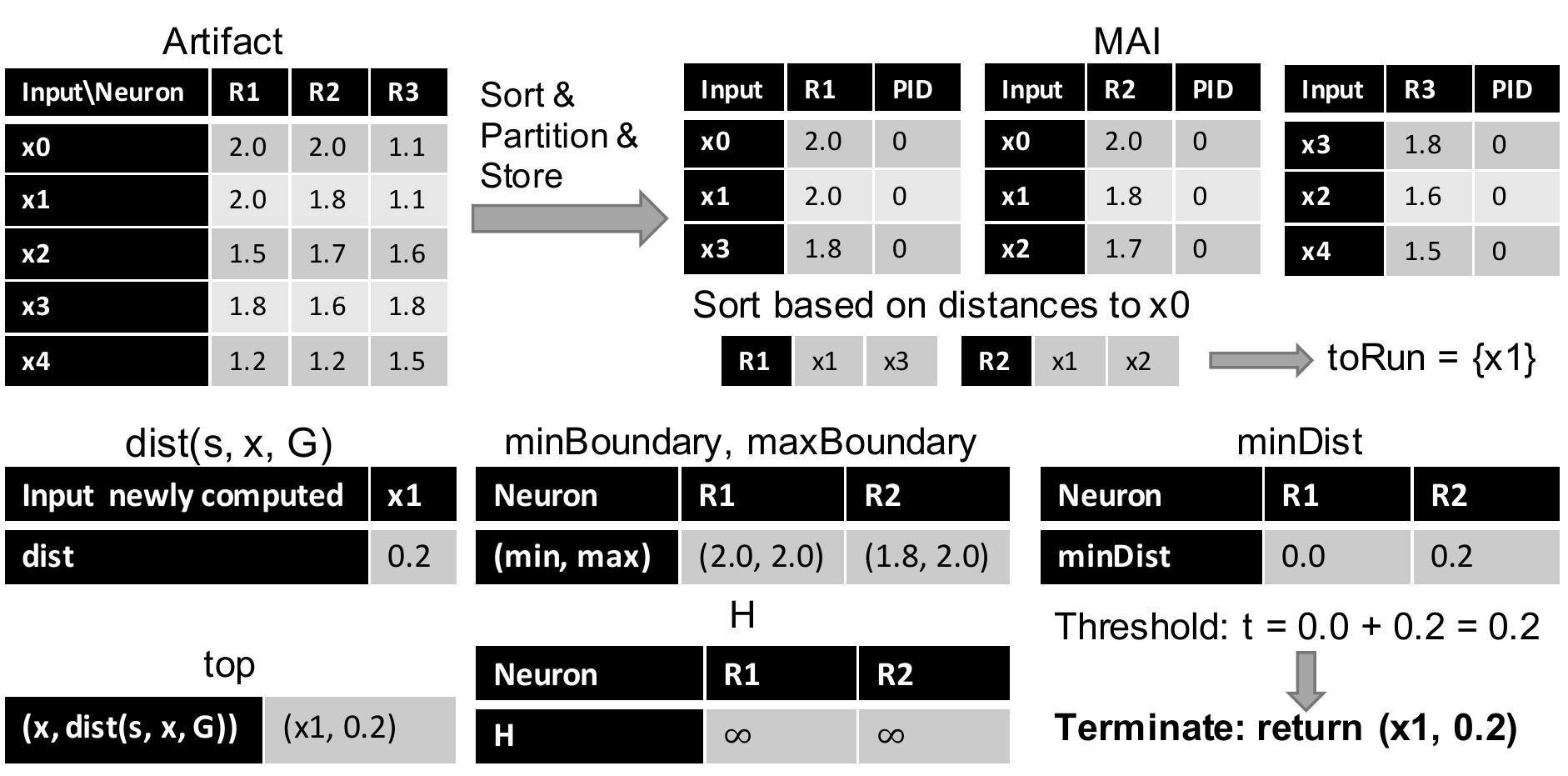}
        \vspace{-2em}
        \caption{An example of constructing MAI ($ratio {=} 0.6$) and query execution for \texttt{topk(}$x0, \{R1, R2, R3\}, 1, l1$-\texttt{distance)} ($batchSize {=} 1$). Despite $x0$ only being in MAI for $R1$ and $R2$, DeepEverest leverages MAI to answer the query after only running DNN inference on $x0$ and $x1$.}
        \vspace{-1.0em}
        \label{fig:exampleMaximumActivationIndex}
    \end{figure}
}

\newcommand{\performance}{
    \begin{figure*}[t!]
        \begin{center}
            \captionsetup{font={color=\revisioncolor}}
            \hspace{0.6em}
            \subfloat{\includegraphics[width=1.0\columnwidth]{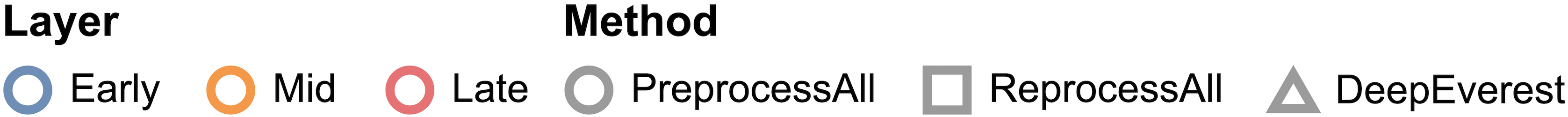}} \hfill%
            \vspace{-1.1em}
            \setcounter{subfigure}{0} 

            \subfloat[\revision{\textit{FireMax, CIFAR10-VGG16}}]{\includegraphics[width=0.33\linewidth]{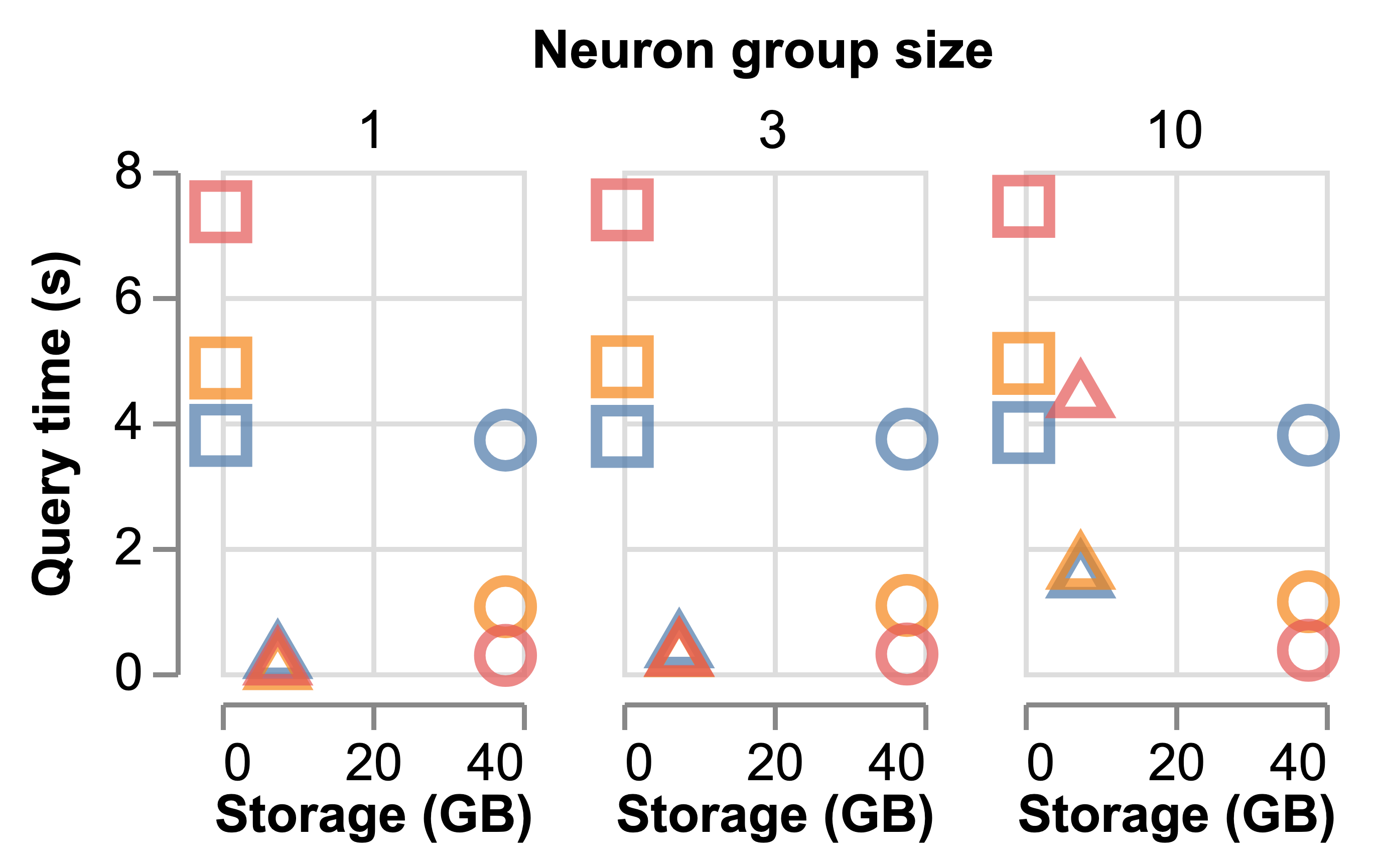}%
                \label{subfig:cifarFireMax}}%
            \hfil
            \subfloat[\revision{\textit{SimTop, CIFAR10-VGG16}}]{\includegraphics[width=0.33\linewidth]{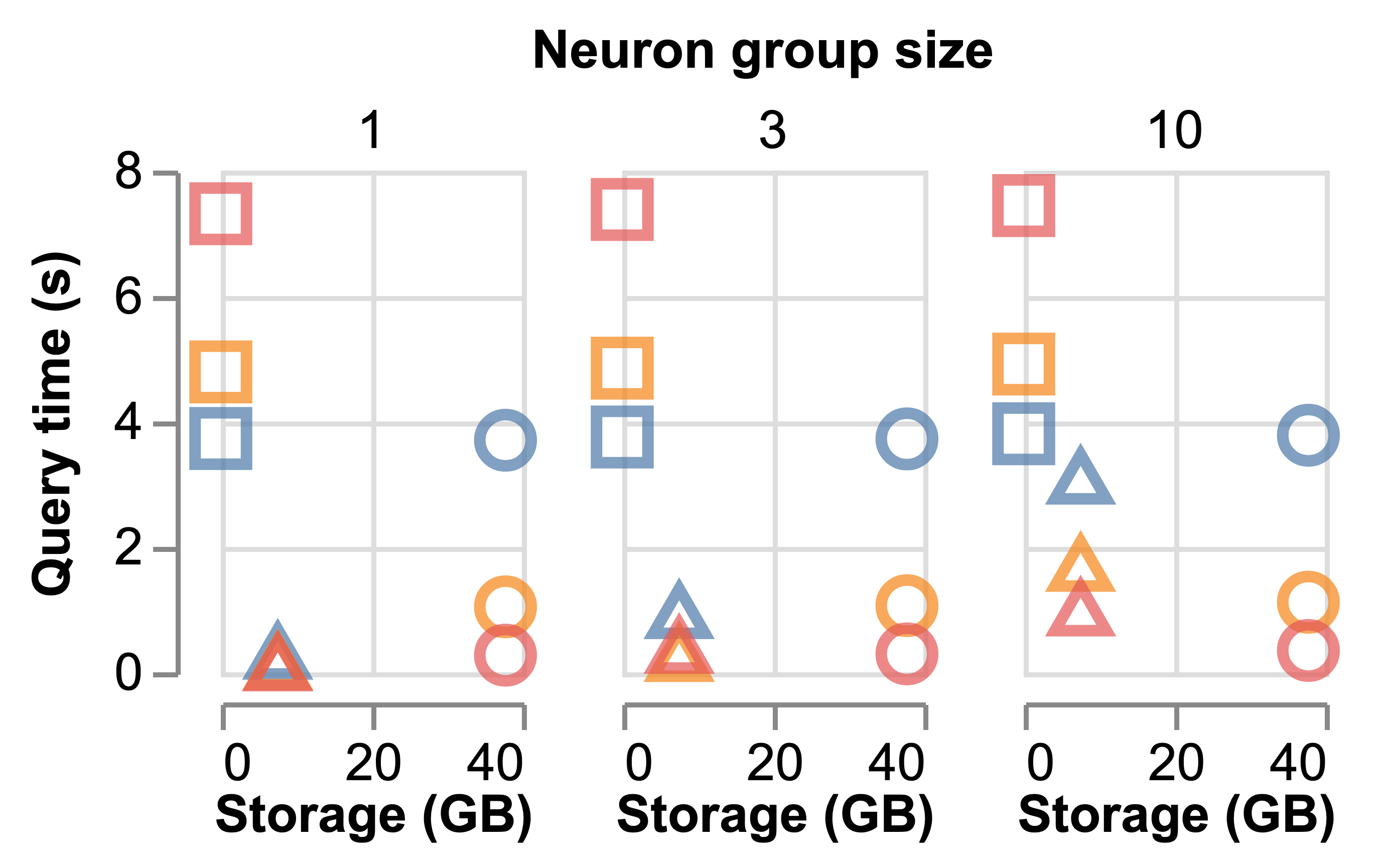}%
                \label{subfig:cifarSimTop}}%
            \hfil
            \subfloat[\revision{\textit{SimHigh, CIFAR10-VGG16}}]{\includegraphics[width=0.33\linewidth]{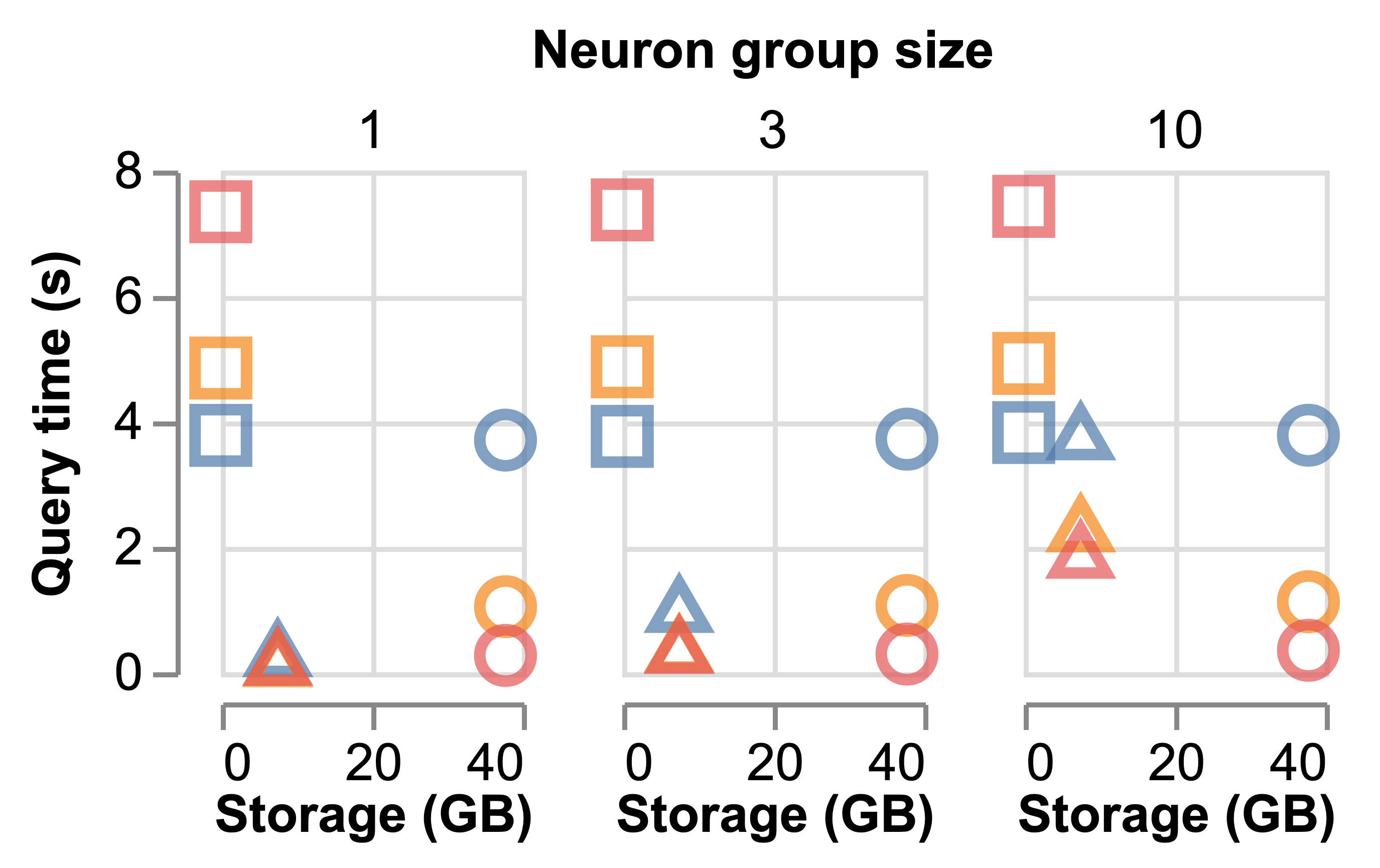}%
                \label{subfig:cifarSimHigh}}%
        \end{center}
        \vspace{-1.2em}
        \begin{center}
            \subfloat[\revision{\textit{FireMax, ImageNet-ResNet50}}]{\includegraphics[width=0.33\linewidth]{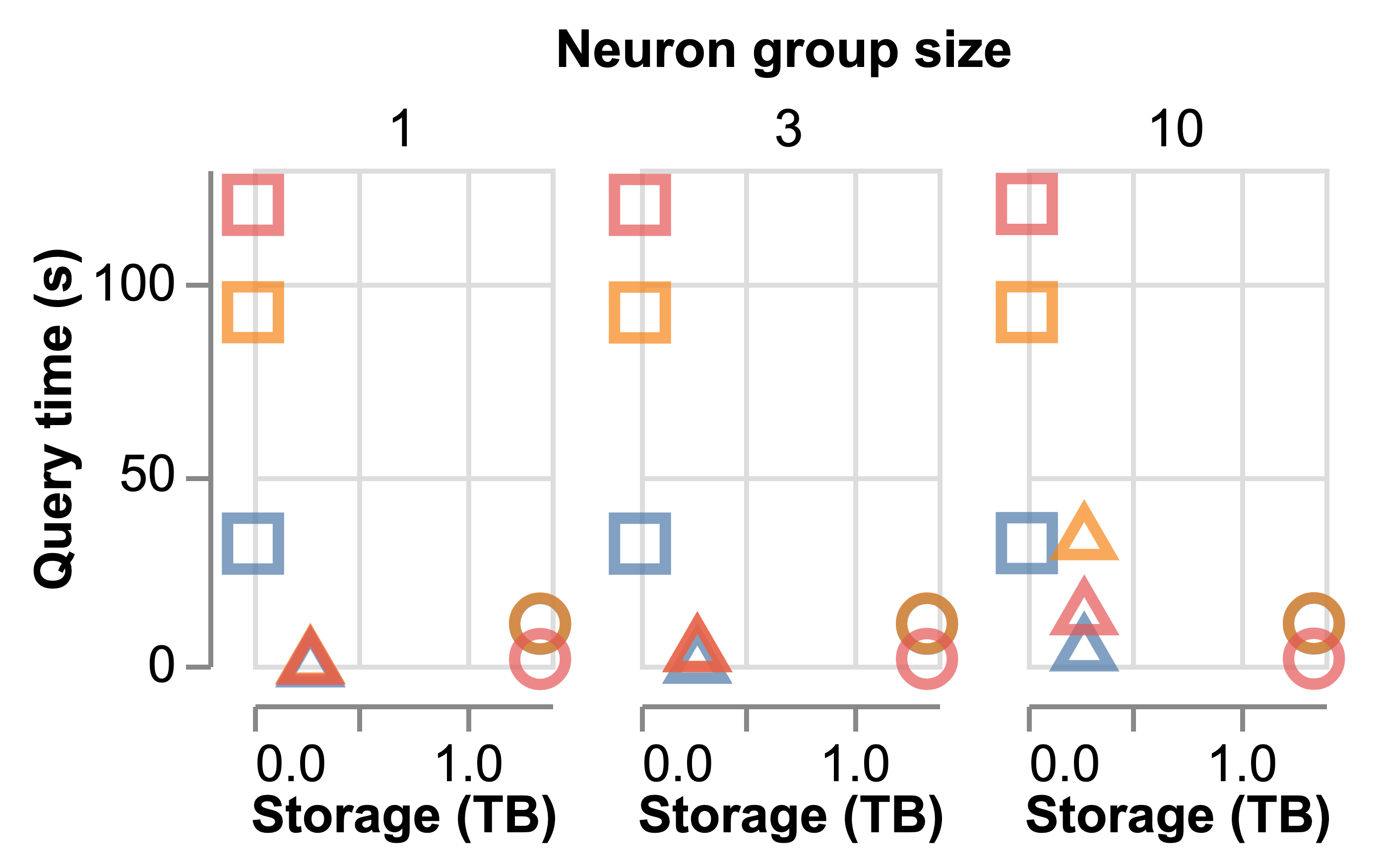}%
                \label{subfig:imagenetFireMax}}%
            \hfil
            \subfloat[\revision{\textit{SimTop, ImageNet-ResNet50}}]{\includegraphics[width=0.33\linewidth]{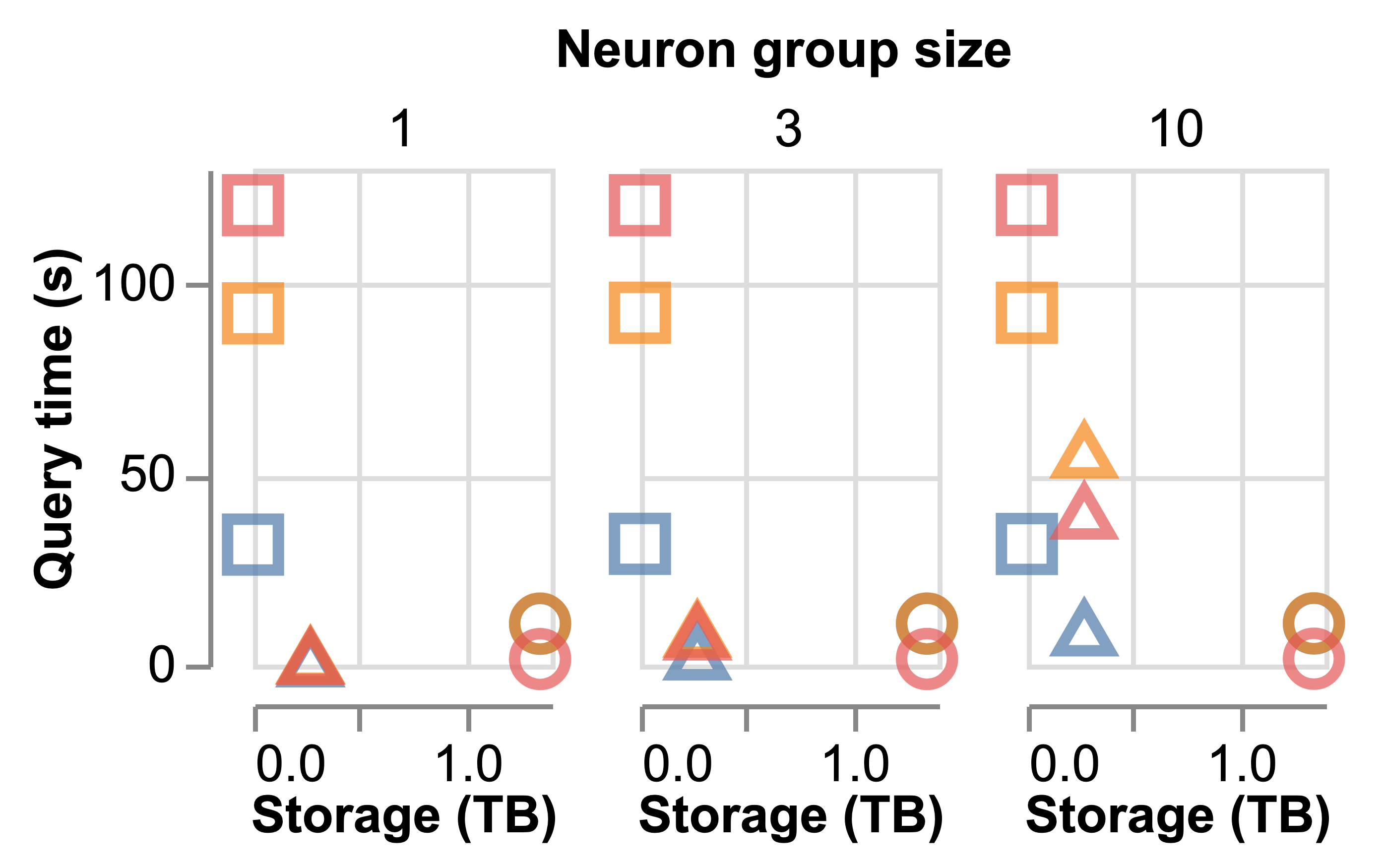}%
                \label{subfig:imagenetSimTop}}%
            \hfil
            \subfloat[\revision{\textit{SimHigh, ImageNet-ResNet50}}]{\includegraphics[width=0.33\linewidth]{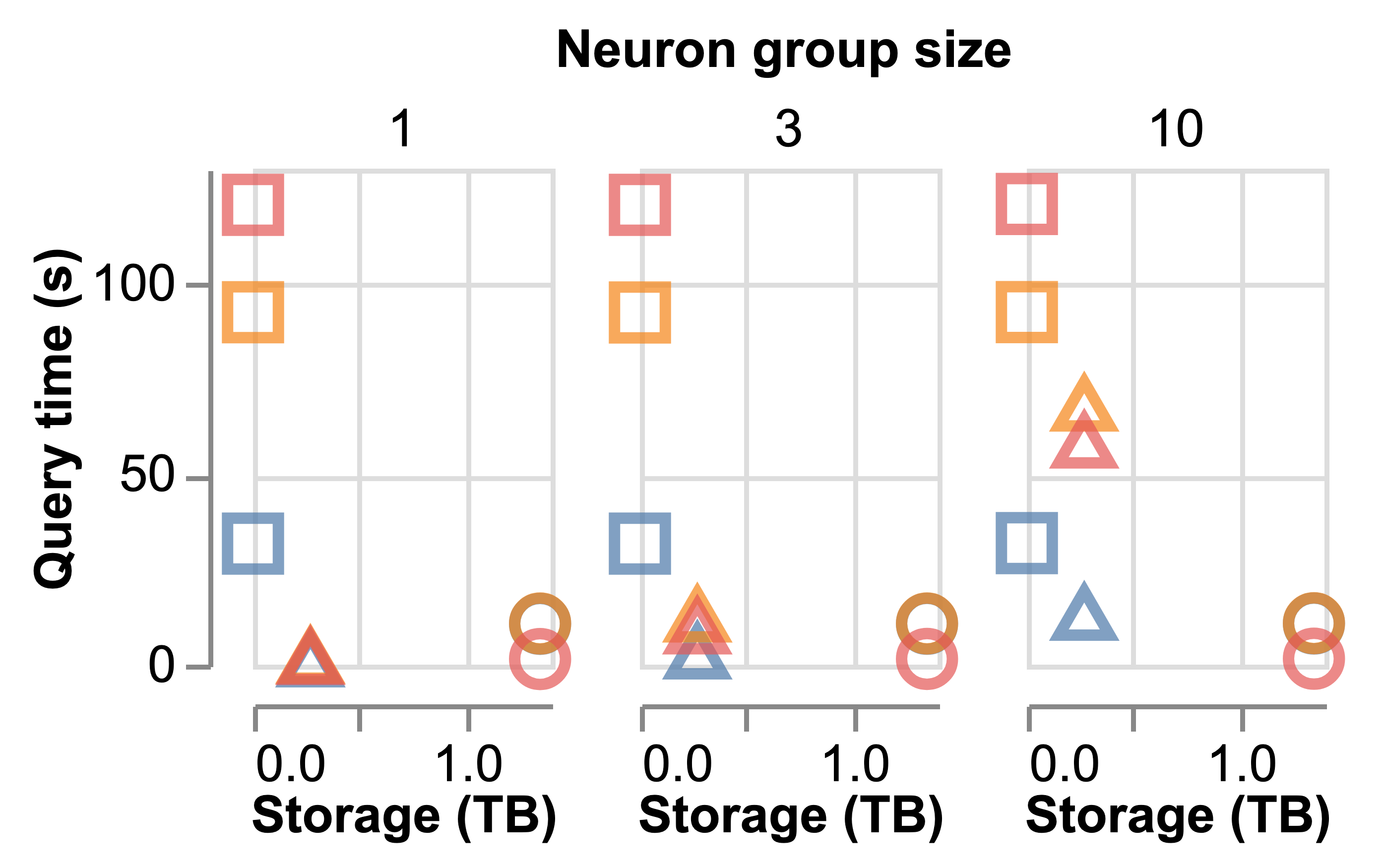}%
                \label{subfig:imagenetSimHigh}}%
        \end{center}
        \vspace{-1.1em}
        \caption{End-to-end individual query times and storage sizes on \textit{CIFAR10-VGG16} and \textit{ImageNet-ResNet50}. $nPartitions$ and $ratio$ of DeepEverest are selected by our \revision{heuristic} algorithm given a storage budget of $20\%$ of full materialization.}
        \vspace{-0.30em}
        \label{fig:performance}
    \end{figure*}
}

\newcommand{\nPartitionsEffect}{
    \begin{figure}[t!]
        \vspace{-0.6em}
        \captionsetup{font={color=\revisioncolor}}
        \hspace{1em}
        \subfloat{\includegraphics[width=0.68\linewidth]{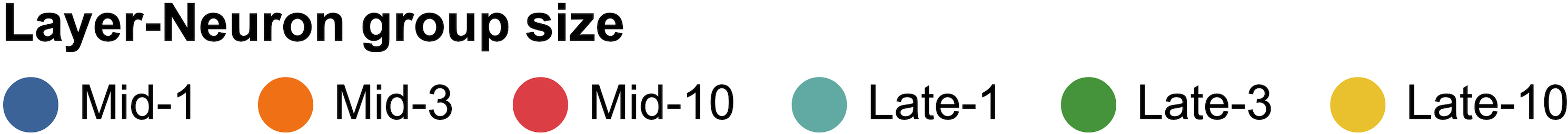}} \hfill%
        \vspace{-1.2em}
        \setcounter{subfigure}{0}

        \subfloat[\textit{CIFAR10-VGG16}]{\includegraphics[width=0.495\linewidth]{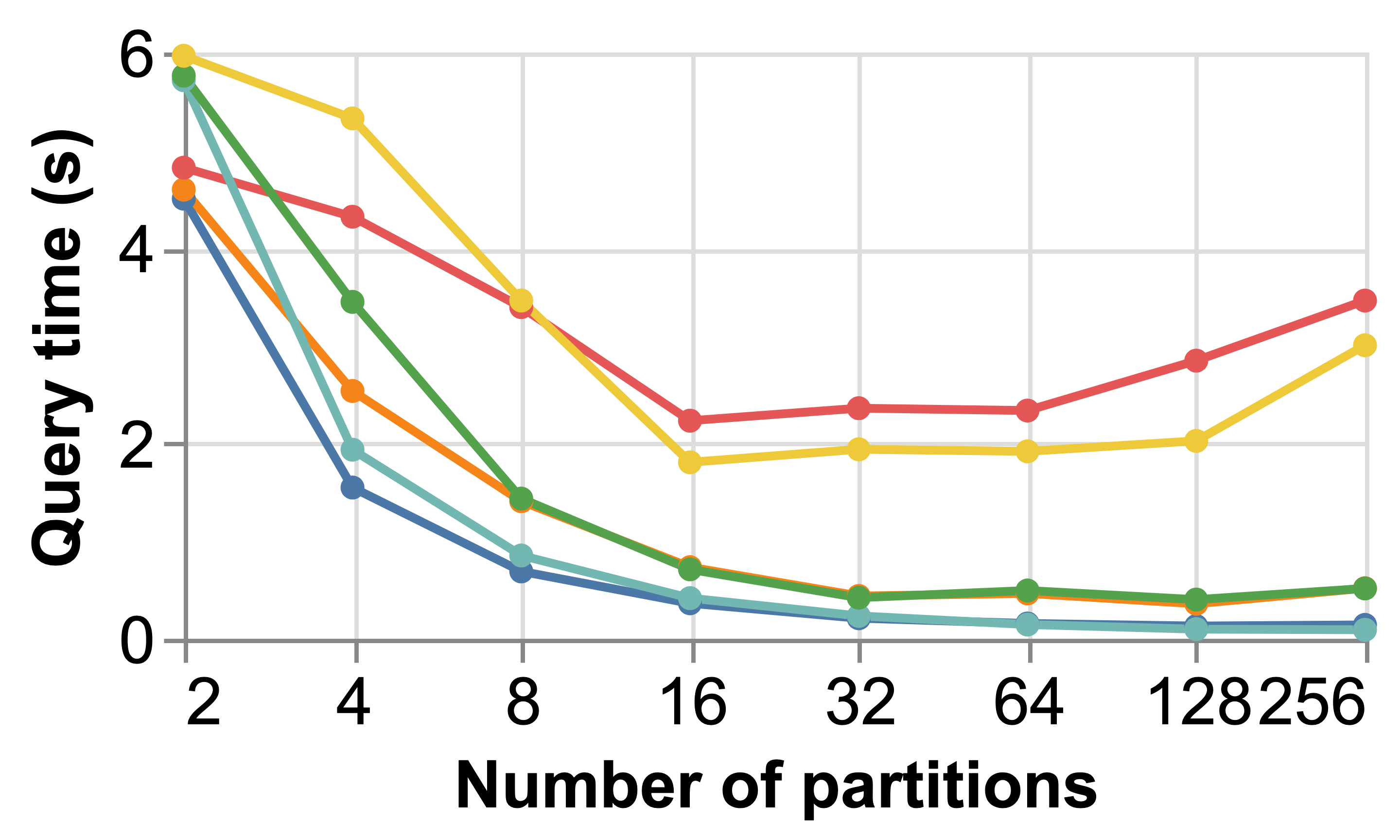}%
                    }%
        \hfil
        \subfloat[\textit{ImageNet-ResNet50}]{\includegraphics[width=0.495\linewidth]{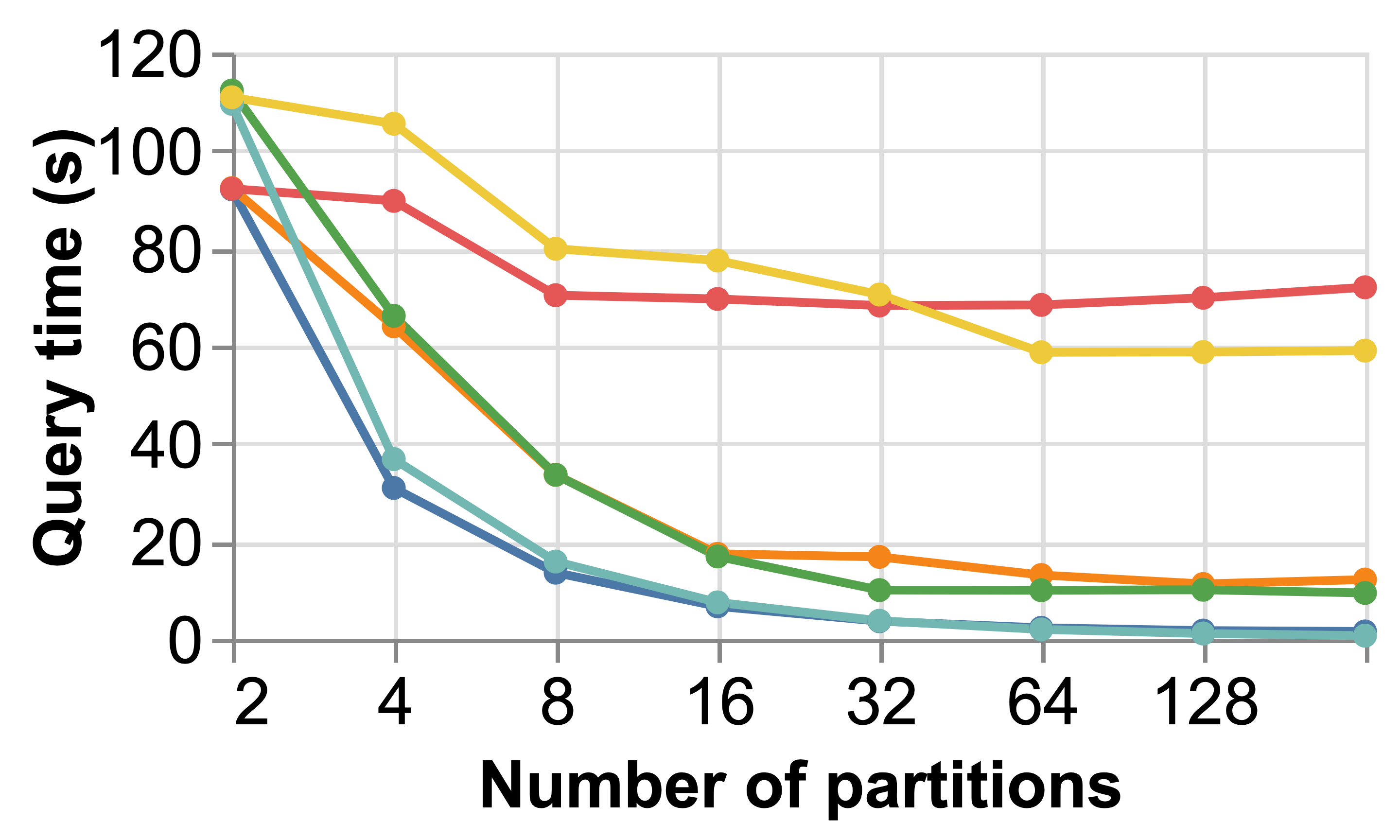}%
                    }%
        \vspace{-1.1em}
        \caption{Query times of \revision{\textit{SimHigh}} queries when varying $nPartitions$. Note the log scale on the $x$-axis.}
        \vspace{-1.8em}
        \label{fig:nPartitionsEffect}
    \end{figure}
}

\newcommand{\ratioEffect}{
    \begin{figure}[t!]
        \captionsetup{font={color=\revisioncolor}}
        \hspace{1em}
        \subfloat{\includegraphics[width=0.9\linewidth]{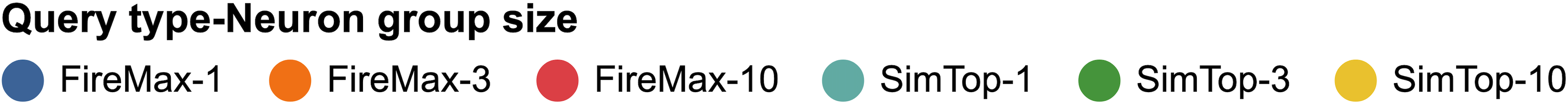}} \hfill%
        \vspace{-1.1em}
        \setcounter{subfigure}{0}

        \subfloat[\textit{CIFAR10-VGG16}]{\includegraphics[width=0.495\linewidth]{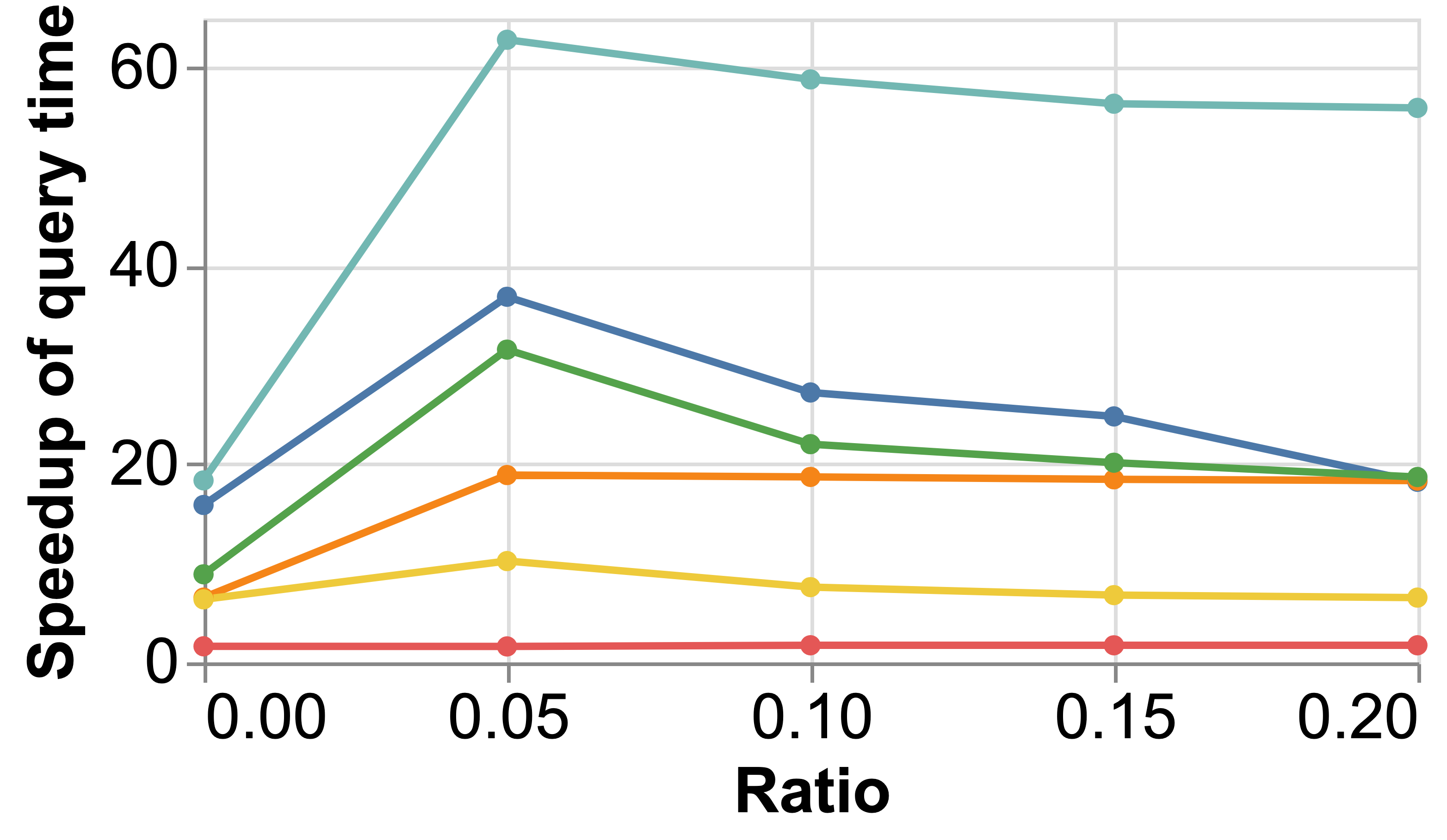}%
                        }%
        \hfil
        \subfloat[\textit{ImageNet-ResNet50}]{\includegraphics[width=0.495\linewidth]{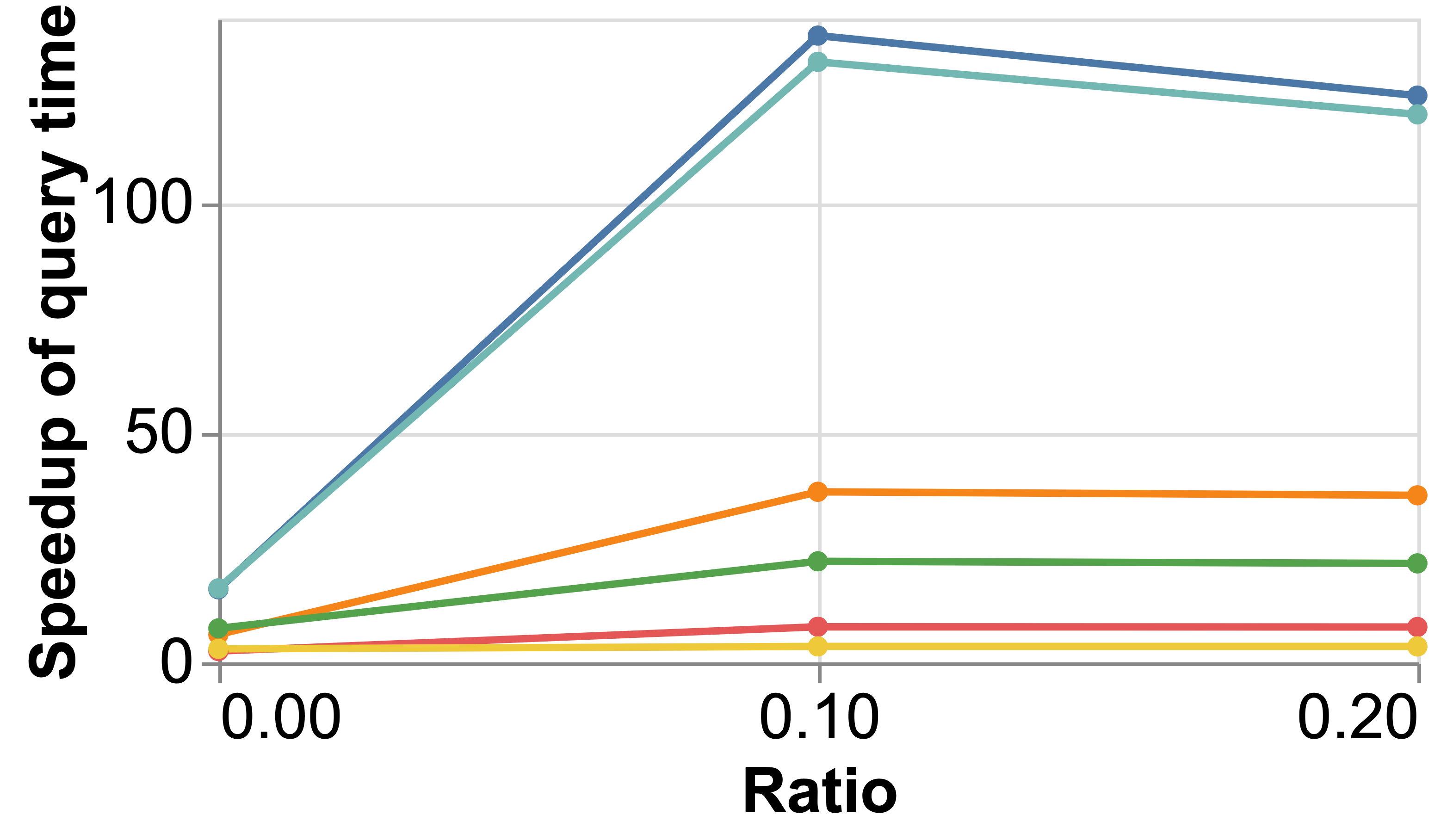}%
                        }%
        \vspace{-1.1em}
        \caption{Speedups of query times (layer: \textit{late}) against \textit{ReprocessAll} when varying $ratio$ with $nPartitions$ set to $16$.}
        \vspace{-1.0em}
        \label{fig:ratioEffect}
    \end{figure}
}

\newcommand{\budgetRobustness}{
    \begin{figure}[t!]
        \captionsetup{font={color=\revisioncolor}}
        \hspace{0.5em}
        \subfloat{\includegraphics[width=0.87\linewidth]{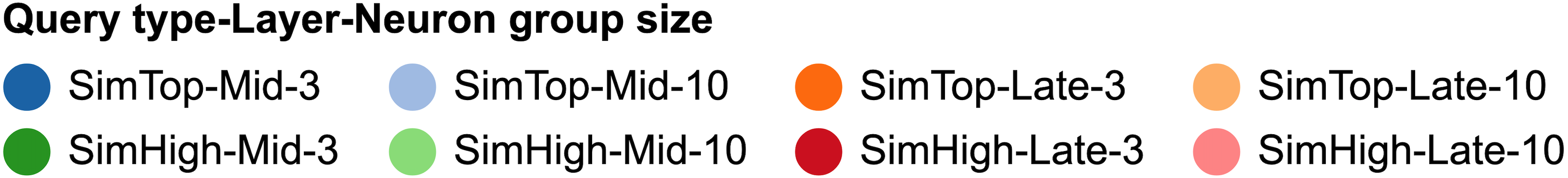}} \hfill%
        \vspace{-1.1em}
        \setcounter{subfigure}{0}

        \subfloat[\textit{CIFAR10-VGG16}]{\includegraphics[width=0.49\linewidth]{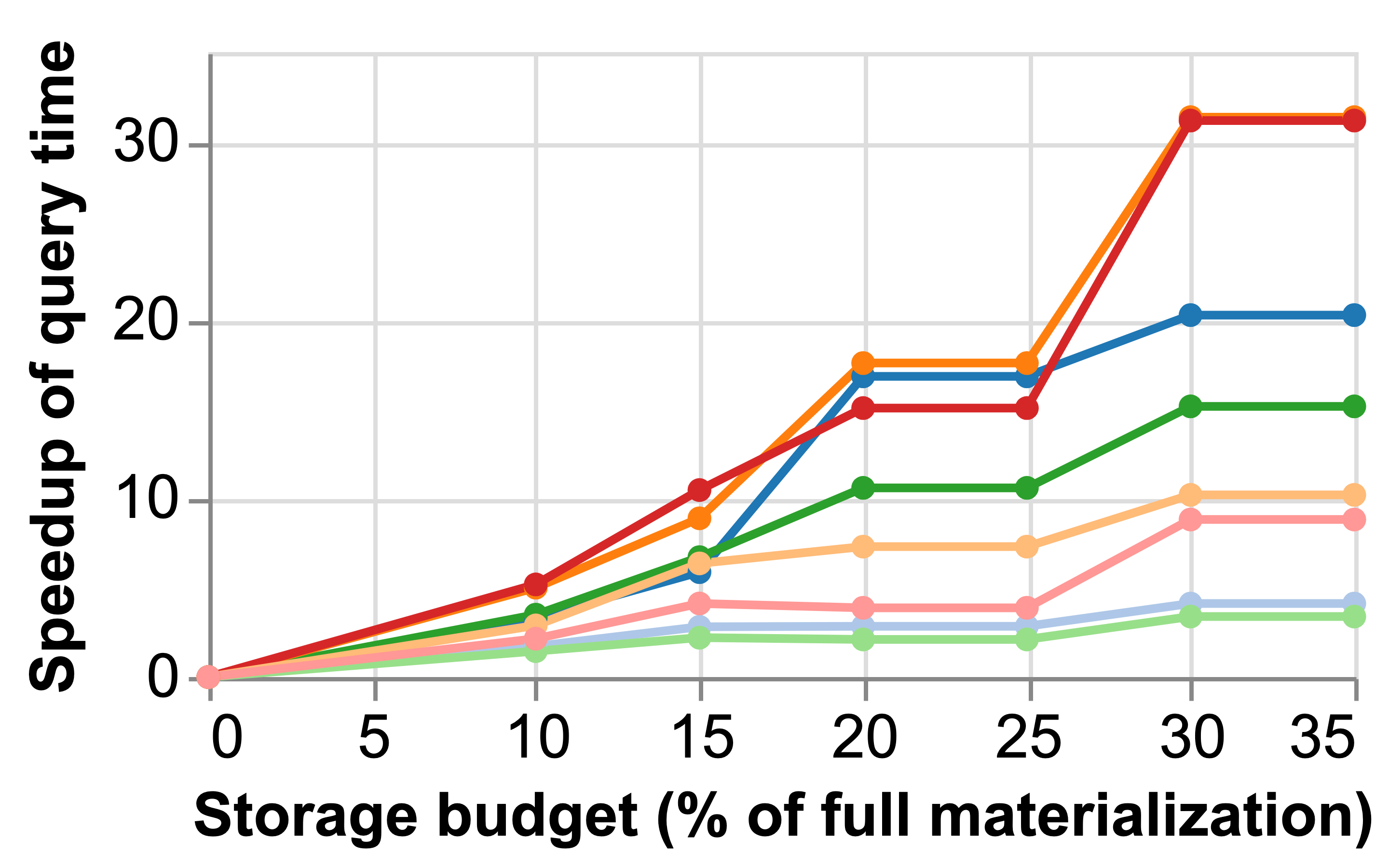}%
            \label{subfig:cifarBudgetRobustness}}%
        \hfil
        \subfloat[\textit{ImageNet-ResNet50}]{\includegraphics[width=0.49\linewidth]{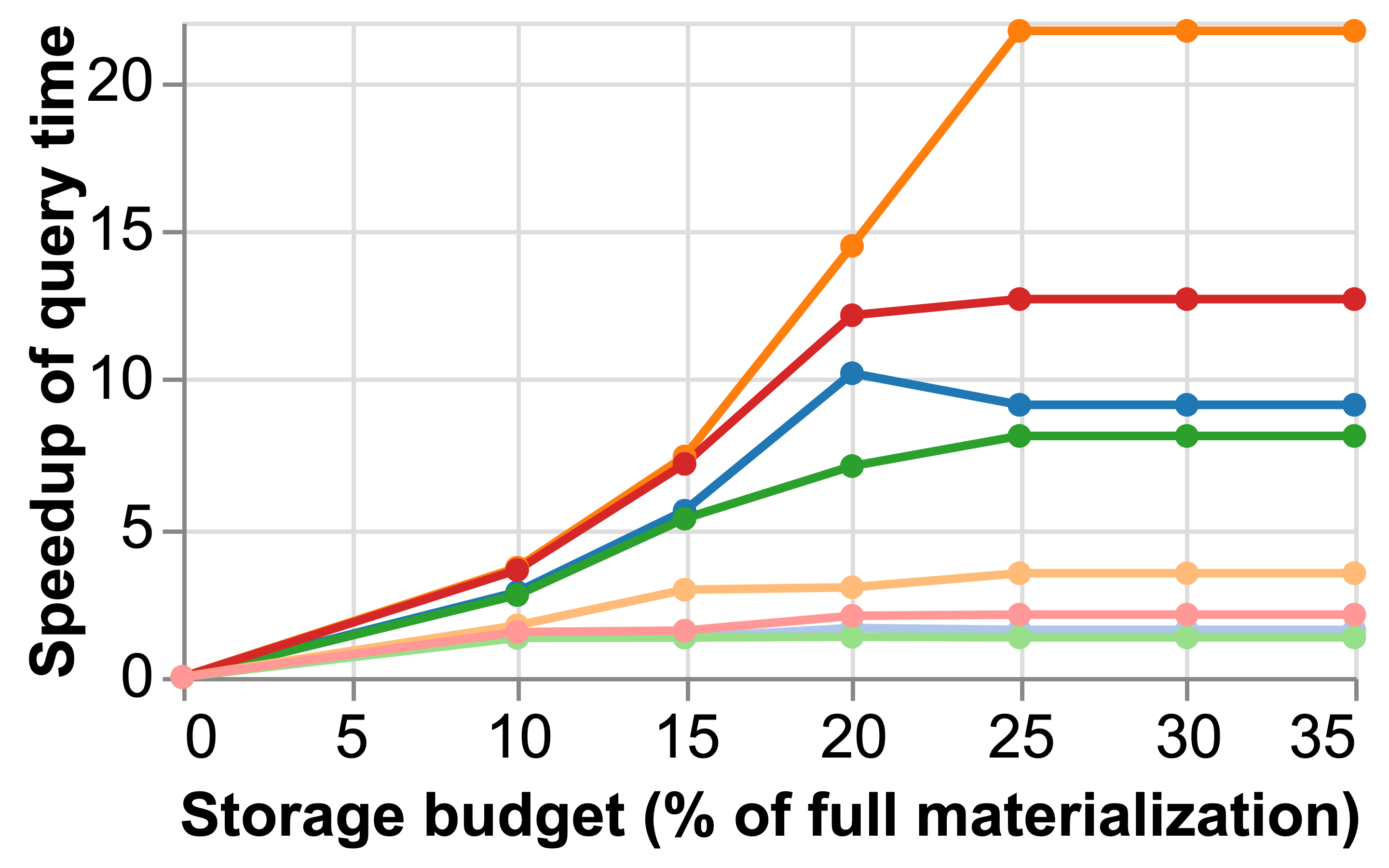}%
            \label{subfig:imagenetBudgetRobustness}}%
        \vspace{-1.1em}
        \caption{Speedups against \textit{ReprocessAll} by DeepEverest when given different storage budgets.}
        \vspace{-0.8em}
        \label{fig:budgetRobustness}
    \end{figure}
}

\newcommand{\CTTIncrementalIdx}{
    \begin{figure*}[t!]
        \captionsetup{font={color=\revisioncolor}}
        \hspace{0.25em}
        \subfloat{\includegraphics[width=0.55\linewidth]{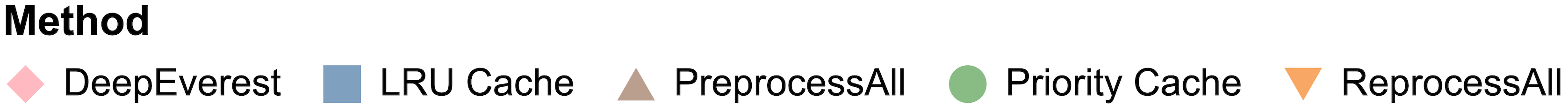}} \hfill%
        \vspace{-1.2em}
        \setcounter{subfigure}{0}
        \begin{center}
            \subfloat[\revision{Workload 1, \cifar}]{\includegraphics[width=0.33\linewidth]{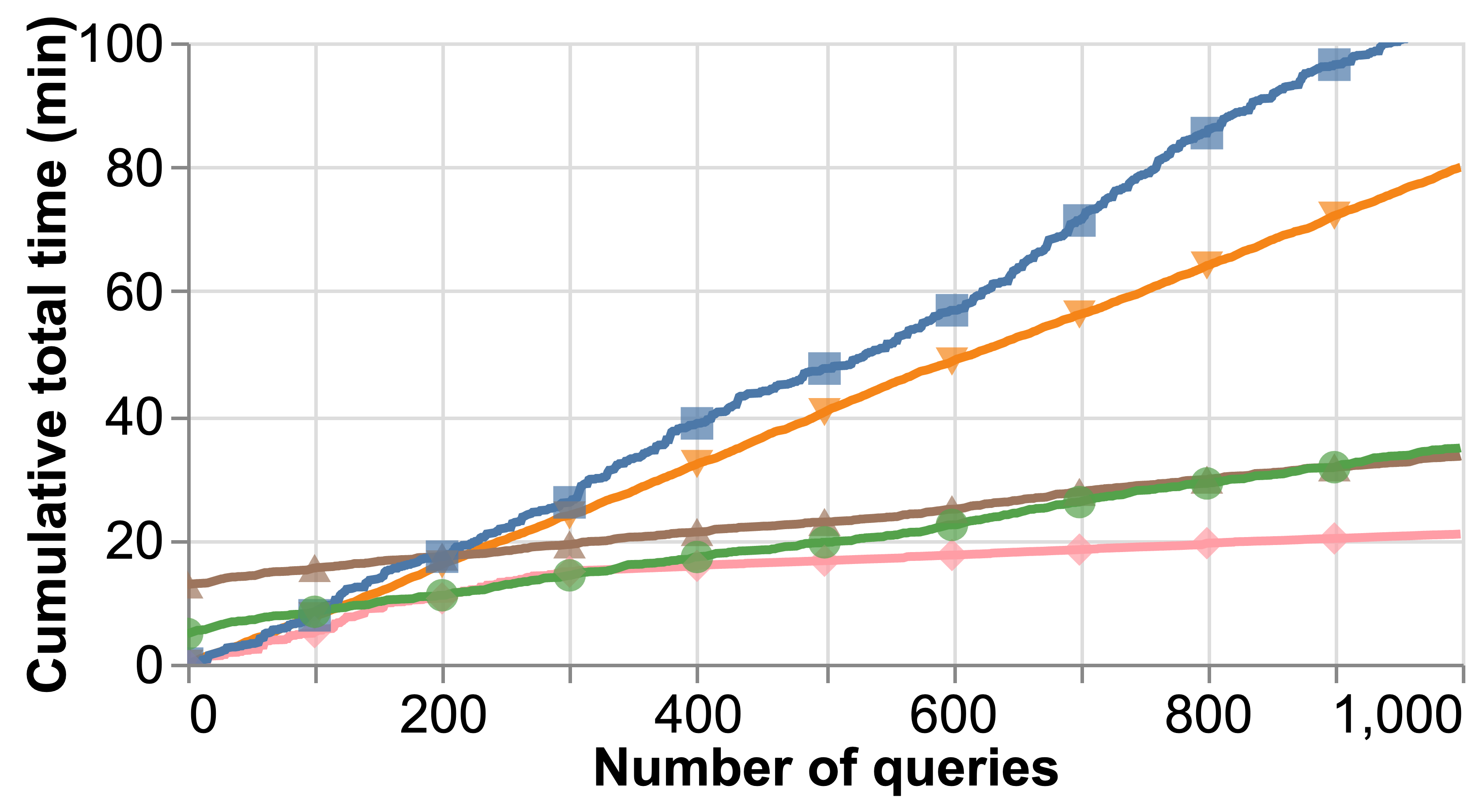}%
                \label{subfig:cifarCTTIncrementalIdxW1}%
            }%
            \hfil
            \subfloat[\revision{Workload 2, \cifar}]{\includegraphics[width=0.33\linewidth]{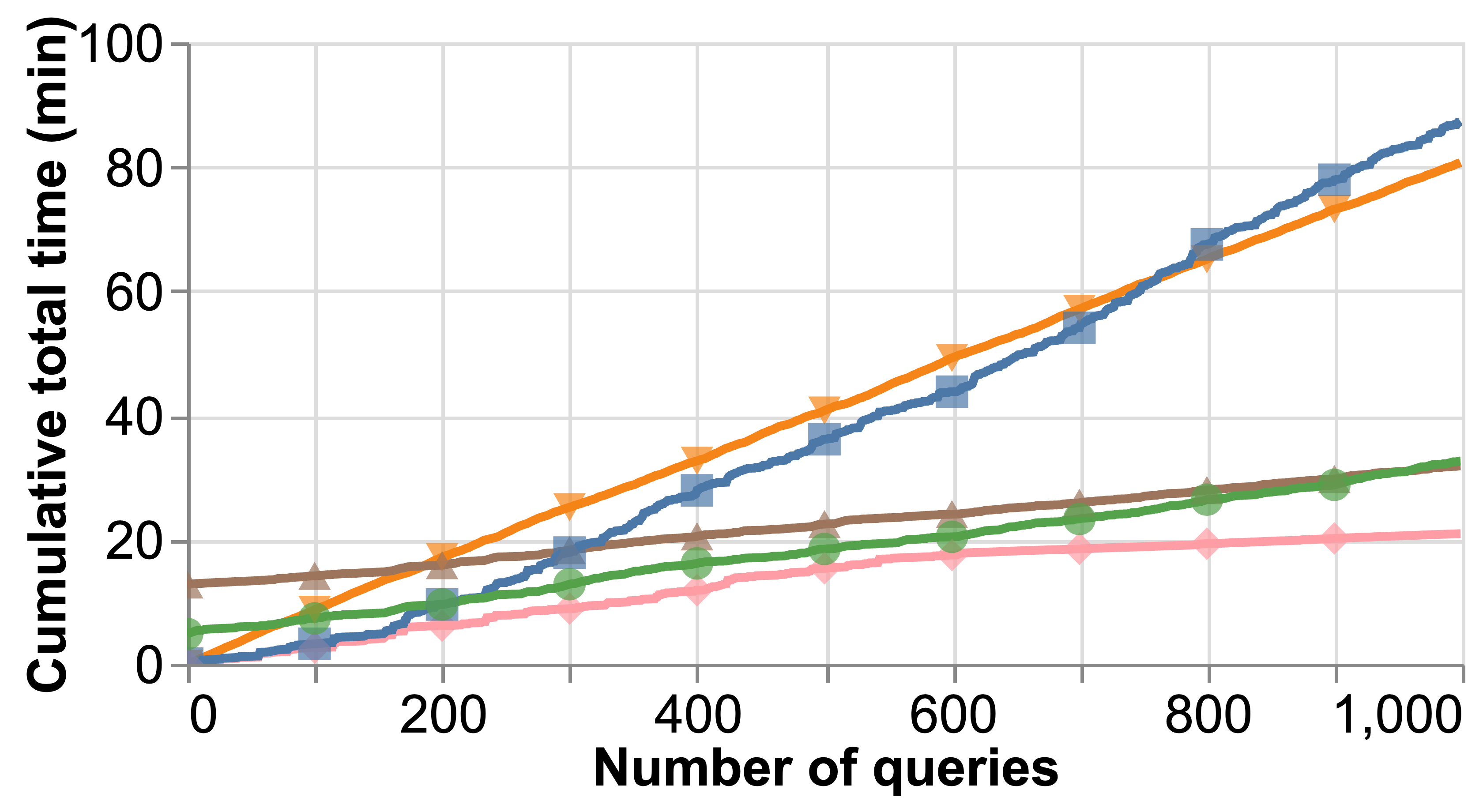}%
                \label{subfig:cifarCTTIncrementalIdxW2}%
            }%
            \hfil
            \subfloat[\revision{Workload 3, \cifar}]{\includegraphics[width=0.33\linewidth]{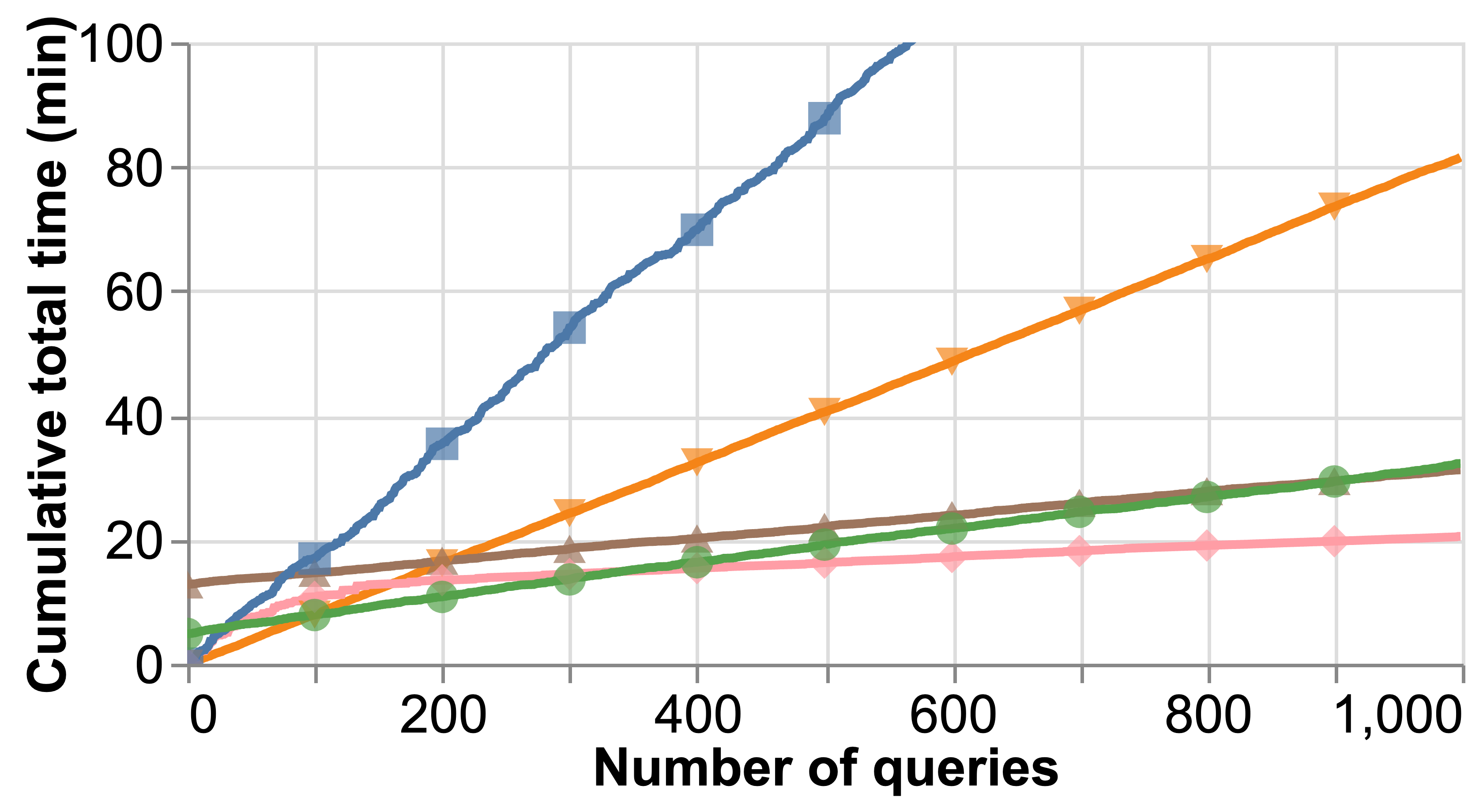}%
                \label{subfig:cifarCTTIncrementalIdxW3}%
            }%
        \end{center}
        \vspace{-0.85em}
        \begin{center}
            \subfloat[\revision{Workload 1, \imagenet}]{\includegraphics[width=0.33\linewidth]{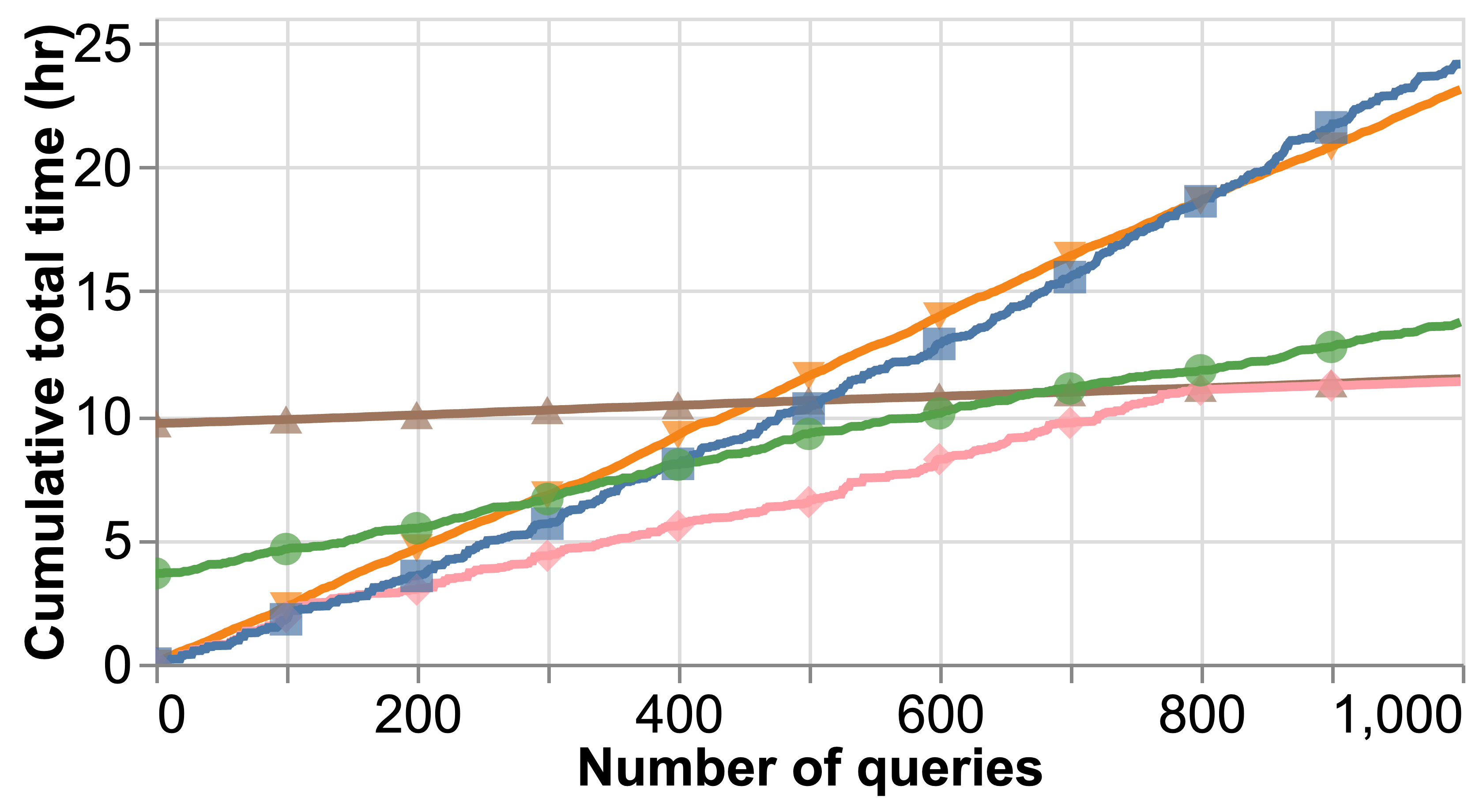}%
                \label{subfig:imagenetCTTIncrementalIdxW1}%
            }%
            \hfil
            \subfloat[\revision{Workload 2, \imagenet}]{\includegraphics[width=0.33\linewidth]{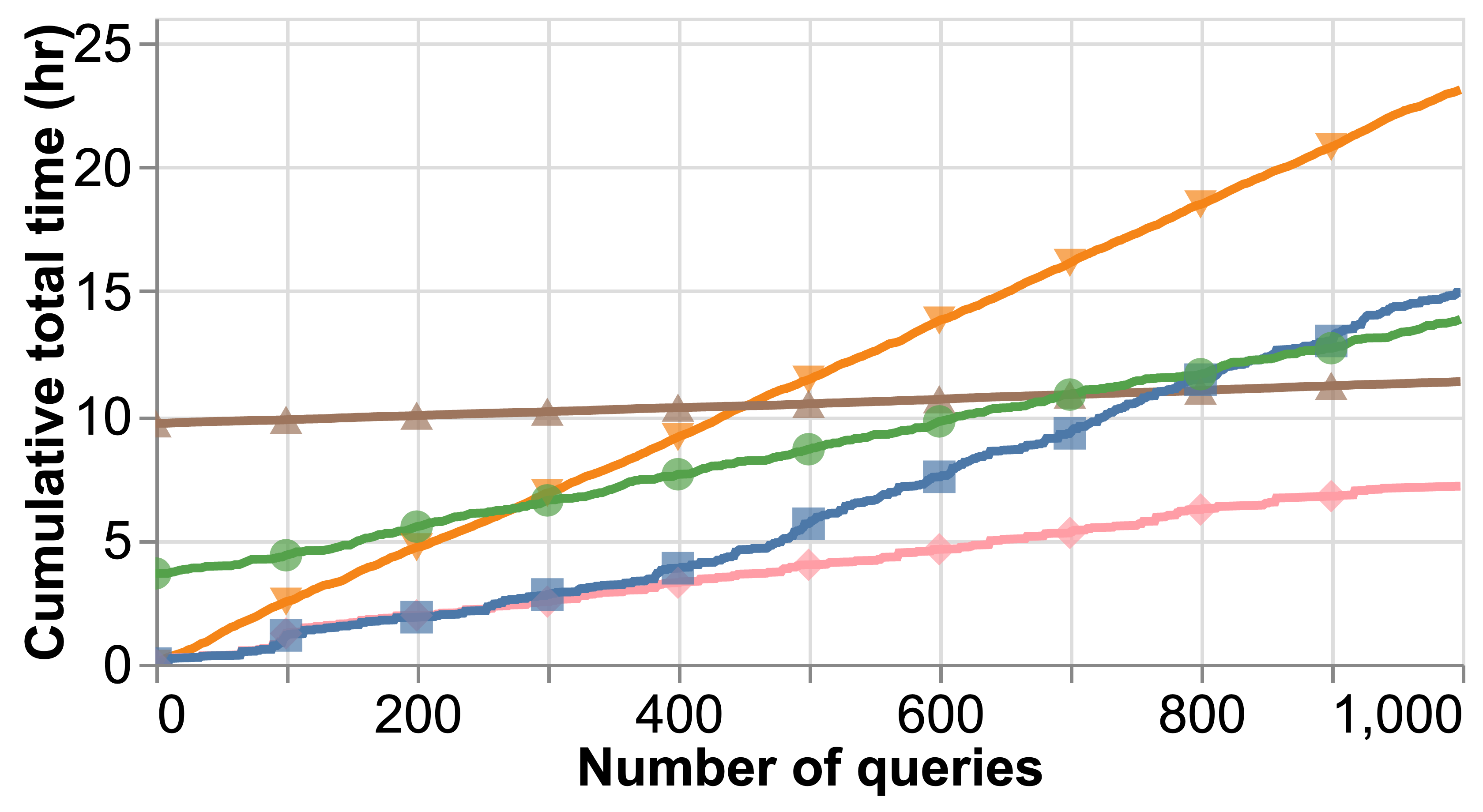}%
                \label{subfig:imagenetCTTIncrementalIdxW2}%
            }%
            \hfil
            \subfloat[\revision{Workload 3, \imagenet}]{\includegraphics[width=0.33\linewidth]{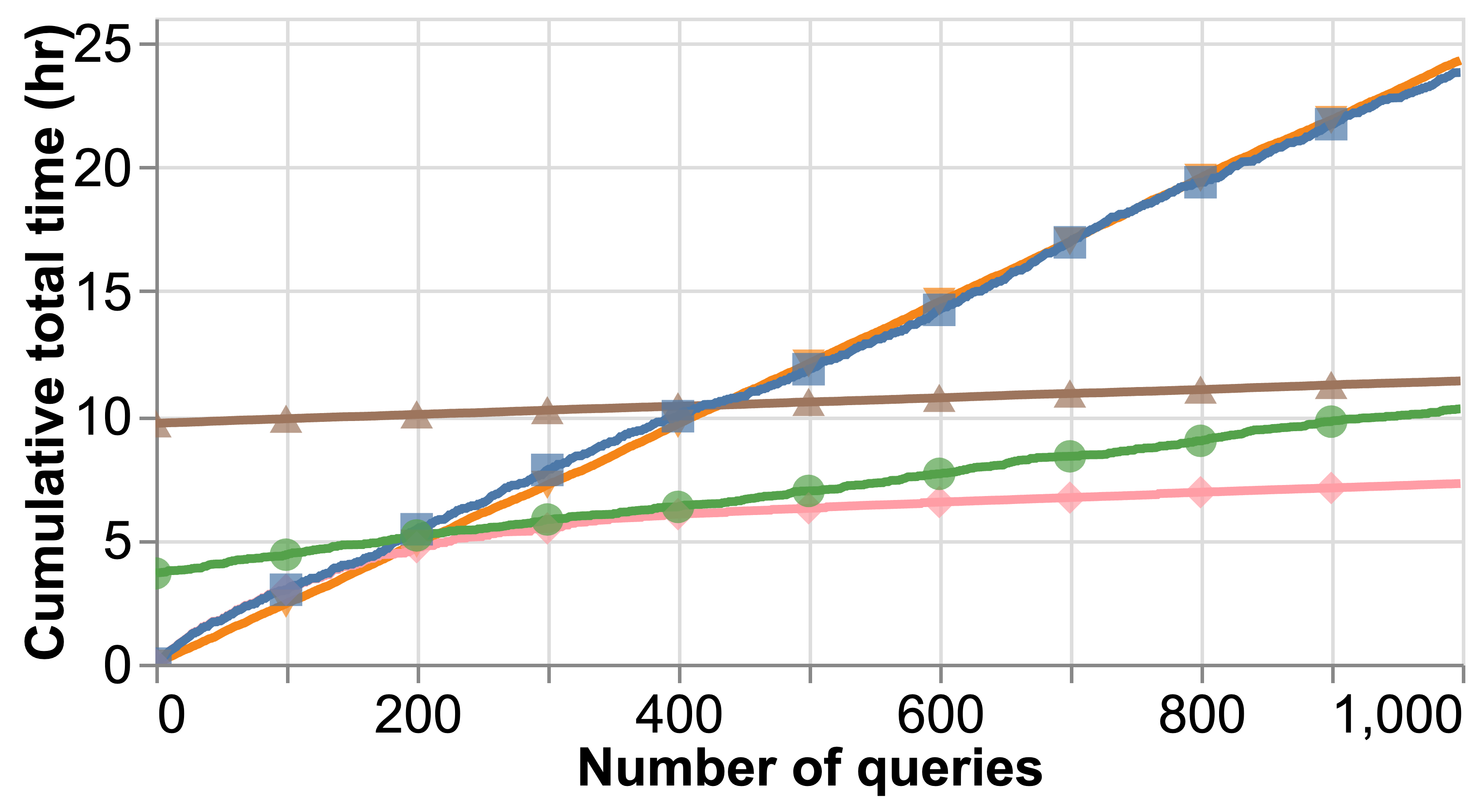}%
                \label{subfig:imagenetCTTIncrementalIdxW3}%
            }%
        \end{center}
        \vspace{-1.08em}
        \caption{Cumulative total time (preprocessing time plus query execution time) for various multi-query workloads.}
        \vspace{-0.070em}
        \label{fig:CTTIncrementalIdx}
    \end{figure*}
}

\newcommand{\preprocessingTime}{
    \begin{figure}[t!]
        \captionsetup{font={color=\revisioncolor}}
        \hspace{0.5em}
        \subfloat{\includegraphics[width=1.0\linewidth]{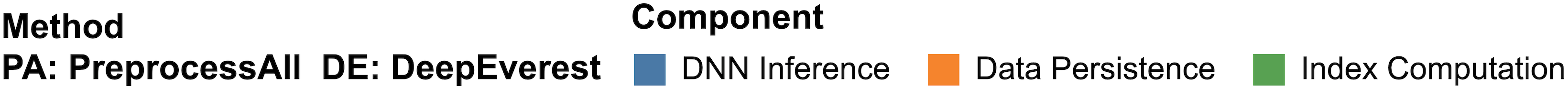}} \hfill%
        \vspace{-2em}
        \setcounter{subfigure}{0}

        \subfloat[\textit{CIFAR10-VGG16}]{\includegraphics[width=0.495\linewidth]{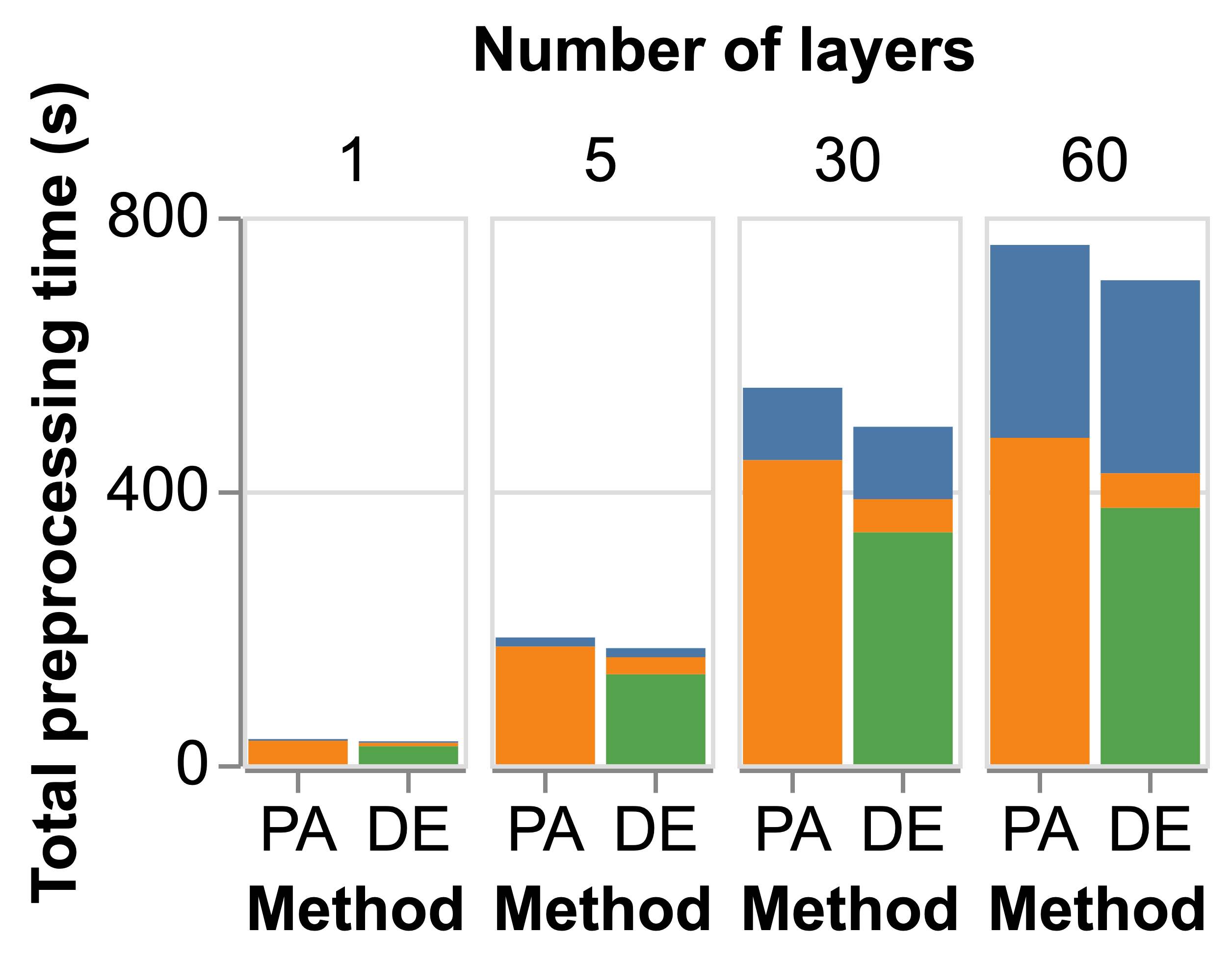}%
                        }%
        \hfil
        \subfloat[\textit{ImageNet-ResNet50}]{\includegraphics[width=0.495\linewidth]{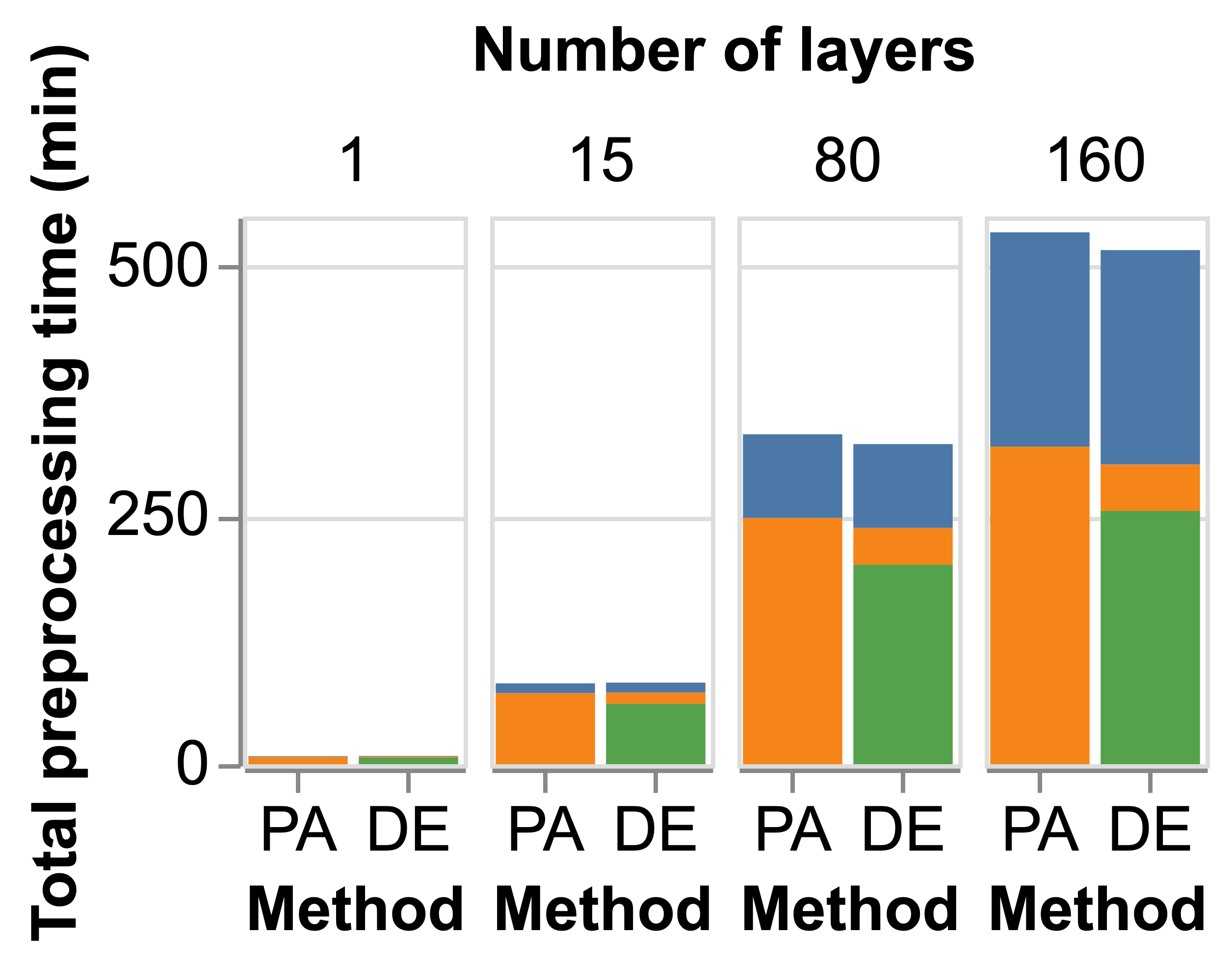}%
                        }%
        \vspace{-1.1em}
        \caption{Cumulative preprocessing times from the first layer to the last layer for \textit{PreprocessAll} and DeepEverest.}
        \vspace{-1.7em}
        \label{fig:preprocessingTime}
    \end{figure}
}

\newcommand{\cifarInterQueryAcc}{
    \begin{figure}[t!]
        \captionsetup{font={color=\revisioncolor}}
        \hspace{0.25em}
        \subfloat{\includegraphics[width=0.3\linewidth]{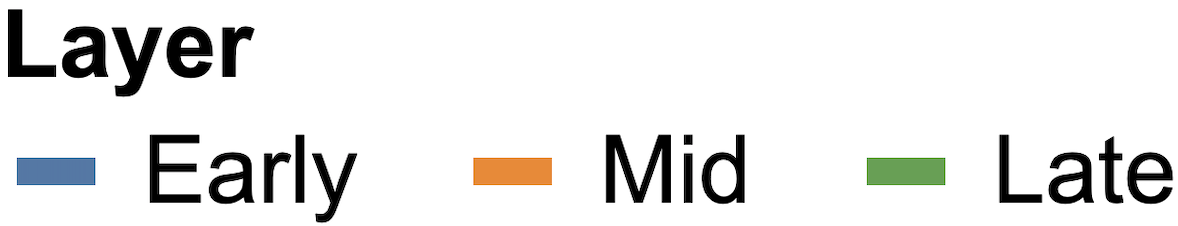}} \hfill%
        \setcounter{subfigure}{0}

        \vspace{-1.1em}
        \subfloat[Sequence 1]{\includegraphics[width=0.49\linewidth]{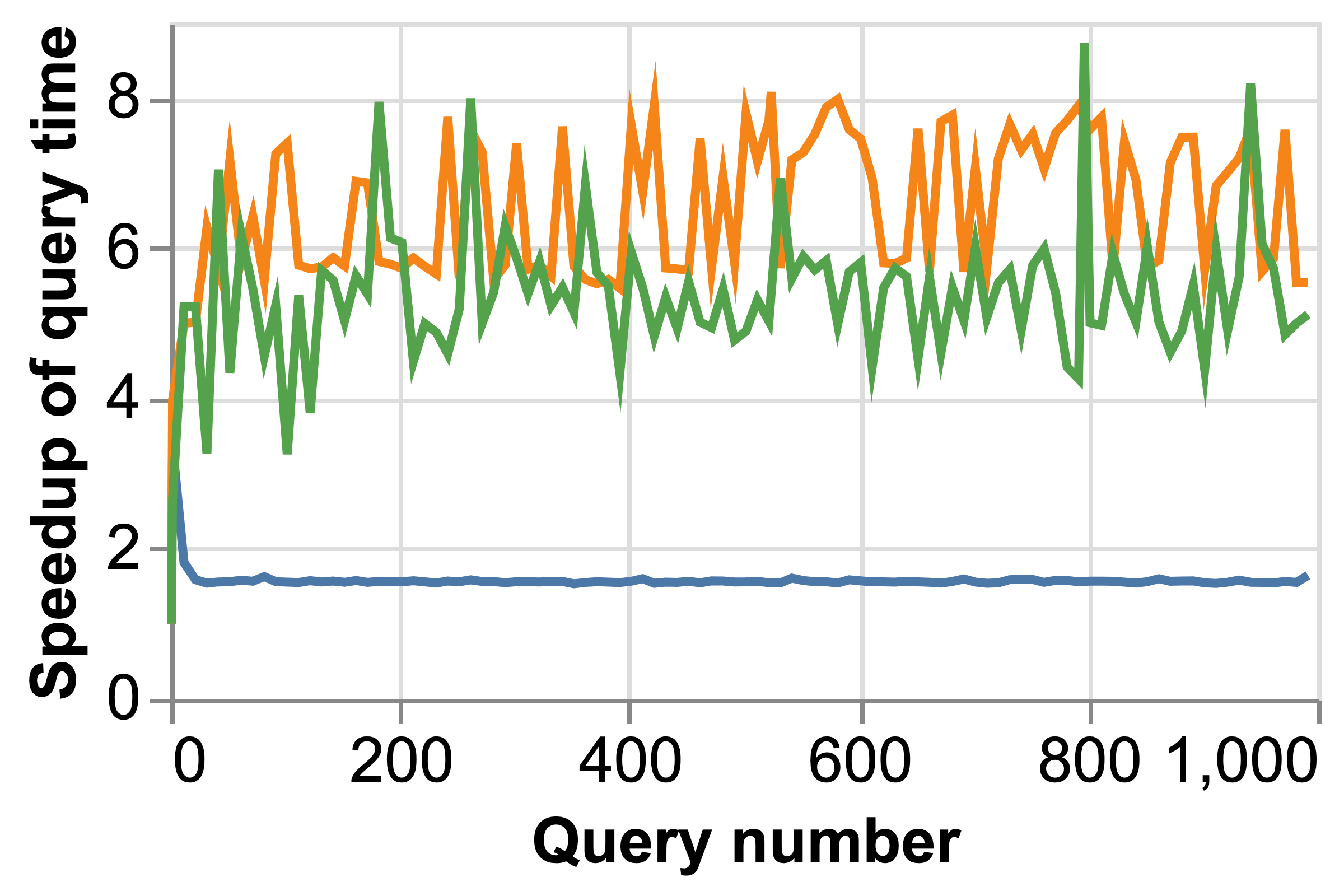}%
            \label{subfig:iqaSequence1}}%
        \hfil
        \subfloat[Sequence 2]{\includegraphics[width=0.49\linewidth]{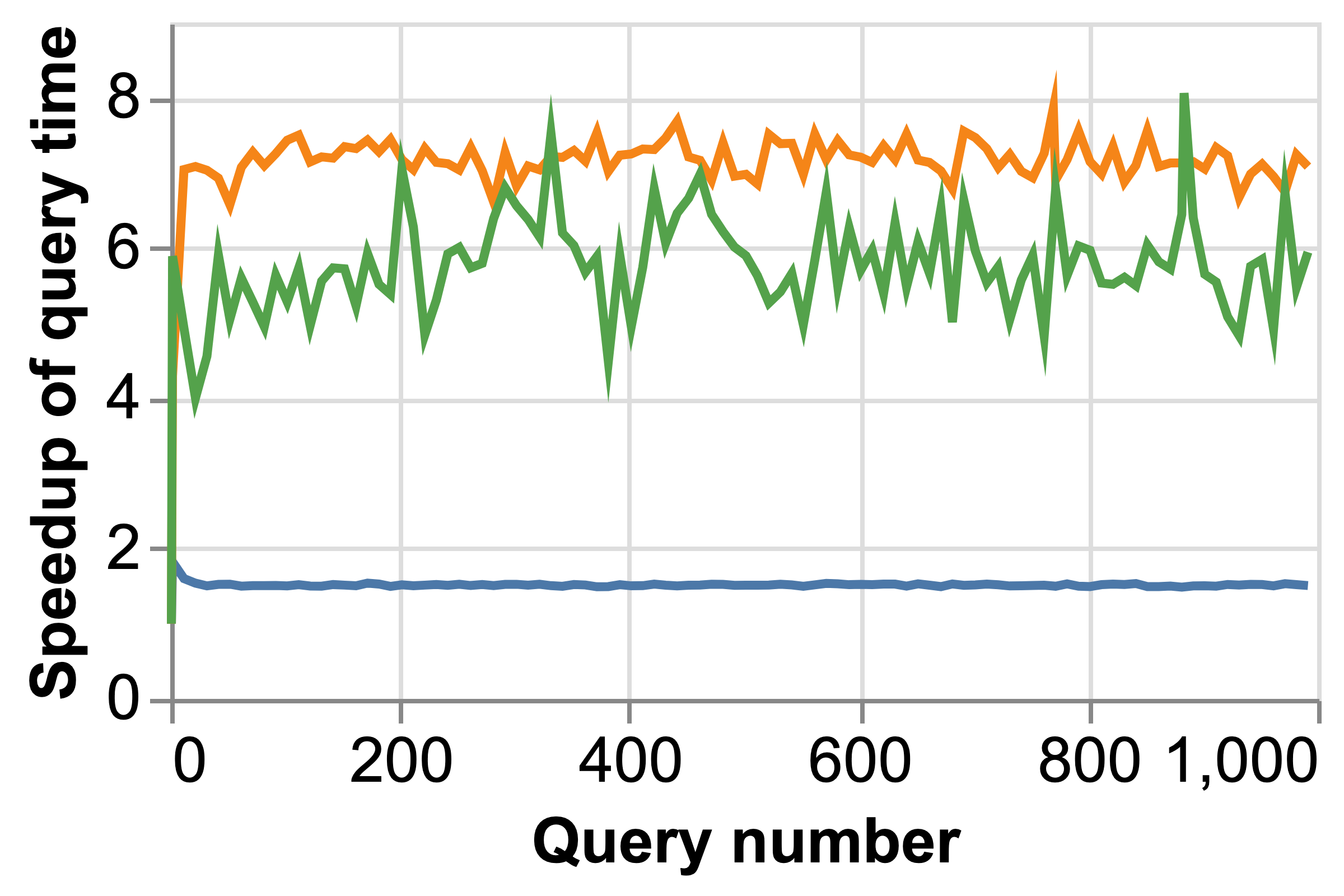}%
            \label{subfig:iqaSequence2}}%
        \vspace{-1.1em}
        \caption{Speedups of query times by DeepEverest with IQA (1 GB cache budget) against DeepEverest without IQA.}
        \vspace{-0.8em}
        \label{fig:cifarInterQueryAcc}
    \end{figure}
}

\newcommand{\dnnBottleneck}{
    \begingroup
    \setlength{\tabcolsep}{2pt}
    \begin{table}[t!]
        \small
        \centering
        \captionsetup{font={color=\revisioncolor}}
        \caption{Query time breakdown for baselines for a \textit{top-$k$ most-similar} query on \imagenet (query: \revision{\textit{SimHigh}}, neuron group size: 3, layer: \textit{late}; detail in \cref{subsec:evaluationSetup}).}
        \vspace{-1.1em}
        \color{\revisioncolor}
        \begin{tabular}{c | c  c  c c}
            \toprule
            Method & \textit{ReprocessAll} & \textit{CTA}~\cite{fagin2003optimal} & \textit{K-D Tree}~\cite{bentley1975multidimensional} & \textit{Ball Tree}~\cite{omohundro1989five} \\
            \midrule
            Total query time & 121.6 s& 121.6 s& 121.8 s& 121.7 s\\
            \midrule
            DNN inference time & 121.4 s& 121.4 s& 121.4 s& 121.4 s\\
            \bottomrule    
        \end{tabular}
        \color{black}
        \normalsize
        \vspace{-0.7em}
        \label{tab:dnnBottleneck}
    \end{table}
    \endgroup
}

\newcommand\headercell[1]{%
   \smash[b]{\begin{tabular}[t]{@{}c@{}} #1 \end{tabular}}}

\newcommand{\nPartitionsNInputsRun}{
    \begingroup
    \begin{table}[t!]
        \captionsetup{font={color=\revisioncolor}}
        \small
        \centering
        \caption{Number of inputs run by the DNN at query time for \revision{\textit{SimHigh}} queries on \textit{CIFAR10-VGG16}.}
        \vspace{-1.1em}
        \centering
        \color{\revisioncolor}
        \begin{tabular}{@{} *{8}{c} @{}}
            \toprule
            \headercell{Layer-Neuron\\group size} & \multicolumn{7}{c@{}}{Number of partitions}\\
            \cmidrule(l){2-8}
            & 4 & 8 & 16 & 32 & 64 & 128 & 256 \\ 
            \midrule
            mid-1 & 3334 & 1429 & 667 & 323 & 159 & 79 & 40 \\ 
            mid-3 & 5462 & 2902 & 1441 & 736 & 727 & 390 & 390 \\ 
            mid-10 & 8941 & 6869 & 4339 & 4215 & 3515 & 3492 & 3316 \\ 
            late-1 & 3334 & 1429 & 667 & 323 & 159 & 79 & 40 \\ 
            late-3 & 5968 & 2372 & 1106 & 618 & 618 & 388 & 391 \\ 
            late-10 & 9008 & 5565 & 2870 & 2745 & 2227 & 1956 & 1919 \\ 
            \bottomrule
        \end{tabular}
        \color{black}
        \normalsize
        \vspace{-0.25em}
        \label{tab:nPartitionsNInputsRun}
    \end{table}
    \endgroup
}

\newcommand{\notationTable}{
    \begingroup
    \setlength{\tabcolsep}{2pt}
    \begin{table}[t!]
        \captionsetup{font={color=\revisioncolor}}
        \small
        \centering
        \caption{Summary of frequently used notation.}
        \vspace{-1.1em}
        \begin{tabular}{ l l }
            \toprule
            Symbol & Meaning \\
            \midrule
            $s$ & Sample input (or target input) from $D$ \\
            \minor{$G$} & Group of neurons from $N$ \\
            \texttt{topk(}$s, G, k,$\textsc{dist}\texttt{)} & \textit{Top-$k$ most-similar} query \\
            $\textsc{dist}$ & Function to compute distances between inputs \\
            \minor{$g_i$ (or $i$ when clear)} & The $i$-th neuron in $G$ \\
            $\texttt{act}(i, x)$ & Activation value of \minor{$g_i$} for input $x$ \\
            $dist(s, x, G)$ & Distance between $s$ and $x$ based on $G$ \\
            $top$ & Set of top-$k$ inputs that are closest to $s$ \\
            \minor{$P_n$} & \minor{Set of partitions for neuron $n$} \\
            \revision{\texttt{getInputIDs}$(n, p)$} & \revision{\texttt{inputID}s of a single partition $p {\in} P_n$} \\
            \revision{\texttt{sPID}$(n)$} & \revision{\texttt{PID} to which $s$ belongs for neuron $n$} \\
            \minor{\texttt{lBnd}$(n, p)$} & Lower bound of a single partition \minor{$p {\in} P_n$} \\
            \revision{\texttt{uBnd}$(n, p)$} & \revision{Upper bound of a single partition $p {\in} P_n$} \\
            \bottomrule
        \end{tabular}
        \normalsize
        \vspace{-0.5em}
        \label{tab:notation}
    \end{table}
    \endgroup
}

\newcommand{\neuralThresholdAlgorithm}{
    \begin{algorithm*}[t!]
        \begin{algorithmic}

            \Function{answerQuery}{$model, D, s, G, k, \textsc{dist}$} \Comment{$model$: the DNN, $D$: dataset, $s$: sample image, $G$: neuron group, $k$: number of results to return, \textsc{dist}: function to compute distances between inputs}
                
                \State $layer \gets \Call{getLayer}{G}$
                \State $P, \texttt{lBnd}, \texttt{uBnd} \gets \Call{loadIndex}{layer}$ \Comment{Load indexes}
                
                \ForAll {$g_i \in G$} \Comment{$P_i$ contains the partitions for neuron $g_i$}
                    \State $P_i \gets \Call{getPartitions}{P, g_i}$ 
                \EndFor

                \State $sampleAct \gets \Call{modelInference}{model, layer, s}$ \Comment{Compute the activations for $s$ by DNN inference}
                \State Initialize \texttt{act} to an empty map that contains the activations of the neuron group for accessed inputs
                \ForAll {$g_i \in G$}
                    \State $\texttt{act}(i, s) \gets sampleAct(i)$
                \EndFor

                \ForAll {$g_i \in G$}
                    \State $\texttt{sPID}(i) = P \myarrow \texttt{getPID}(g_i, s)$
                    \State Initialize the list $dPar(i)$
                    \ForAll {$p \in P_i$} \Comment{$dPar(i, p)$: the distance from each partition $p$ for neuron $g_i$ to $s$}
                        \If {$p == \texttt{sPID}(i)$} $dPar(i, p) \gets 0$
                            \ElsIf {$p < \texttt{sPID}(i)$}
                                \text{ }$dPar(i, p) \gets \texttt{lBnd}(i, p) - \texttt{act}(i, s)$ 
                            \Else
                                \text{ }$dPar(i, p) \gets \texttt{act}(i, s) - \texttt{uBnd}(i, p)$ 
                        \EndIf
                    \EndFor
                \EndFor
                
                \ForAll {$g_i \in G$} \Comment{$ord(i)$: the order by which the partitions for neuron $g_i$ are accessed}
                    \State $ord(i) \gets \Call{argsort}{dPar(i)}$ 
                \EndFor
                
                \ForAll {$g_i \in G$} \Comment{Initialization of some variables}
                    \State $F_i \gets 1$, $V_i \gets 1$
                    \State $minBoundary_i \gets \infty$, $maxBoundary_i \gets -\infty$
                \EndFor
                \State $c \gets 0$, $top \gets \emptyset$ \Comment{Starting with $c = 0$; $top$: current top-$k$ result set}
                \State $inputRun \gets \{s\}$ \Comment{$inputRun$: set of inputs that have been run for DNN inference}

                \While {\textbf{True}}
                    \ForAll {$g_i \in G$}
                        \If {$ord(i, c)$ does not exist}
                            $\Return \text{ } top$ \Comment{Return if all partitions have been seen}
                        \EndIf
                        \State $toRun_i \gets P \myarrow \texttt{getInputIDs}(i, ord(i, c))$ 
                    \EndFor

                    \If {$exitFlag$}
                        \textbf{break}
                    \EndIf

                    \State $toRunUnion \gets \bigcup_{g_i \in G} toRun_i \setminus inputRun$
                    \State $toRunAct \gets \Call{modelInference}{model, layer, toRunUnion}$ \Comment{Run DNN inference in batches}
                    
                    \ForAll {$x \in toRunUnion$}
                        \State Initialize the list $diff$
                        \ForAll {$g_i \in G$}
                            \State $\texttt{act}(i, x) \gets toRunAct_{i, x}$
                            \State $diff_i \gets |\texttt{act}(i, x) - \texttt{act}(i, s)|$
                        \EndFor
                        \State $dist(s, x, G) \gets \textsc{dist}(diff)$ \Comment{Compute the distance between $x$ and $s$}
                        \If {$|top| < k$ \textbf{or} $dist(s, x, G) < \Call{getMaxDist}{top}$}
                            \State $\Call{update}{top, x, dist(s, x, G)}$ \Comment{Update $top$ if $x$ is one of the $k$-most similar seen}
                        \EndIf
                    \EndFor

                    \ForAll {$g_i \in G$}
                        \ForAll {$x \in toRun_i$}
                            \State $minBoundary_i \gets \Call{min}{minBoundary_i, \texttt{act}(i, x)}$
                            \State $maxBoundary_i \gets \Call{max}{maxBoundary_i, \texttt{act}(i, x)}$
                        \EndFor

                        \If {$ord(i, c + 1)$ does not exist}
                            $F_i \gets \infty$
                        \EndIf
                        \If {$ord(i, c) == 0$}
                            $V_i \gets \infty$
                        \EndIf

                        \State $minDist_i \gets \Call{min}{F_{i} \cdot |minBoundary_{i} - \texttt{act}(i, s)|, V_{i} \cdot |maxBoundary_{i} - \texttt{act}(i, s)|}$

                    \EndFor

                    \State $t \gets \textsc{dist}(minDist)$ \Comment{Calculate the threshold $t$}

                    \If {$|top| == k$ \textbf{and} $\Call{getMaxDist}{top} \le t$}
                        \textbf{break} \Comment{Termination condition}
                    \EndIf

                    \State $inputRun \gets inputRun \cup toRunUnion$
                    \State $c \gets c + 1$

                \EndWhile

                \State $\Return \text{ } top$

            \EndFunction

        \end{algorithmic}

        \caption{The Neural Threshold Algorithm for \textit{top-k most-similar} queries.}

        \label{alg:neural-threshold-algorithm}
    \end{algorithm*}
}

\vspace{0.60em}
\section{Introduction} \label{sec:introduction}

Deep neural networks (DNNs) are increasingly used by machine learning (ML) applications.
When training and deploying DNNs, interpretation is important for researchers to understand what their models learn. 
DNN interpretation is a relatively new field of research, and techniques are evolving. 
While many new approaches are developed, they often do not scale with the size of the datasets and models~\cite{sellam2019deepbase}. 
The problem we address in this paper is the efficient execution of a common class of DNN interpretation queries.

The fundamental building blocks of DNN interpretation are neurons. 
Each neuron outputs an activation value as the input is propagated through the network. 
DNN interpretation techniques often perform analysis on these activation values of neurons~\cite{bau2017network, olah2018building, zhou2018revisiting, gonzalez2018semantic, zhou2018interpreting, bau2020understanding}. When DNNs are trained on tasks such as image classification or scene synthesis, there emerge individual neurons and groups of neurons that match specific human-interpretable concepts~\cite{zhou2014object, gonzalez2018semantic, bau2020understanding}, such as ``human faces'' and ``trees''.

To understand what individual neurons and groups of neurons learn and detect, researchers often ask \textit{interpretation by example} queries~\cite{lipton2018mythos}, \revision{which are important constituents of the class of post-hoc interpretation methods that are applied to trained models, as opposed to methods that achieve interpretability by restricting model complexity~\cite{molnar2019}.} 
A widely used \minor{\textit{interpretation by example}} query is, \textit{``find the top-$k$ inputs that produce the highest activation values for an individual neuron or \minor{a group of neurons}''}~\cite{goodfellow2009measuring, zeiler2014visualizing, szegedy2014intriguing, girshick2014rich, zhou2014object, yosinski2015understanding, ma2016going, kahng2017cti}. Another common query is, \textit{``for any input, find the k-nearest neighbors in the dataset using the activation values of a group of neurons based on the proximity in the latent space \minor{defined by the group of neurons}''}~\cite{caruana1999case, mikolov2013distributed, karpathy2015visualizing, amershi2015modeltracker, nguyen2016synthesizing, papernot2018deep, wallace2018interpreting}. 
\revision{
These queries help with investigating and understanding the functionalities of neuron groups by tying those functionalities to the input examples in the dataset. 
Moreover, these queries are staple techniques for verifying hypotheses of what groups of neurons learn and detect. 
For instance, Mikolov et al. ask the latter query to find the nearest neighbors of words in the latent space to examine the learned representations after training the word2vec model~\cite{mikolov2013distributed}.} 
As a concrete example, consider a DNN trained to classify images. A user may be interested in understanding what parts of a dog image cause the DNN to predict its class. The user may inspect the maximally activated neurons of different layers in the DNN based on the conjecture that groups of maximally activated neurons act as semantic detectors of features in the image (e.g., floppy ears). To investigate whether these neurons exhibit similar behavior for other images, the user may then ask for the most similar images to the sample image based on the activation values of a group of neurons.

This paper presents a system called DeepEverest that focuses on accelerating the aforementioned two kinds of queries: (1) find top-$k$ inputs that produce the highest activation values for a user-specified group of neurons, and (2) find the top-$k$ most similar inputs based on a given input's activation values for a user-specified neuron group. A group of neurons consists of one or more neurons within a layer of the DNN. We call the first type of query the \textit{top-$k$ highest} query and the second type of query the \textit{top-$k$ most-similar} query.

Executing these \textit{interpretation by example} queries efficiently with low storage overhead is challenging. 
A baseline approach is to materialize the activation values for all inputs and all neurons. However, this approach requires significant storage space. For example, storing all the activations uncompressed of ResNet50 for 10,000 images occupies 1.35~TB of disk storage. 
At the other extreme, computing all activation values at query time imposes no storage overhead, but is compute-intensive and extremely slow because it requires DNN inference to compute the activation values on the entire dataset at query time. For instance, answering a \textit{top-$k$ most-similar} query that targets a relatively late layer of ResNet50 on a dataset of 10,000 images takes more than 120 seconds, which renders the DNN interpretation process tedious.

Further, although the target query is a $k$-nearest neighbor (KNN) search, existing approaches that accelerate KNN queries are not applicable.  KNN methods rely on building efficient data structures such as trees~\cite{bentley1975multidimensional, omohundro1989five, liu2006new} or hash tables~\cite{datar2004locality, andoni2006near} in advance for faster query execution later.  One could try to build a single, large, multidimensional data structure for all neurons in each layer. However, such an index would not perform well because of its very large dimensionality. DNNs frequently have layers with multiple thousands of neurons, thus dimensions. One could build data structures for all possible neuron groups that a user could query. However, this would either limit the user to a small set of possible queries or would be prohibitively expensive both in time and storage because the number of possible neuron groups grows exponentially with the number of neurons in each layer. Additionally, in all cases, precomputing and storing all activation values in such data structures would add prohibitive storage overhead.

While many systems have been developed to enable various forms of DNN interpretation~\cite{amershi2015modeltracker, kahng2016visual, kahng2017cti, krishnan2017palm, ribeiro2016should, sellam2019deepbase}, 
none supports flexible and efficient \textit{interpretation by example} queries. 
In prior work~\cite{DBLP:conf/icde/MehtaPBC20}, we investigated the use of sampling for model diagnosis. 
That work, however, focused only on aggregate queries. 
The closest work to DeepEverest is MISTIQUE~\cite{vartak2018mistique}. It introduces storage techniques such as compression and quantization, which are orthogonal to DeepEverest and could complement our approach. It is, however, possible to use some of MISTIQUE's techniques as a caching algorithm, which we compare against in our experiments.

In DeepEverest, we design an index called \minor{the Neural Partition Index (NPI)}, and an efficient query execution algorithm, called \minor{the Neural Threshold Algorithm (NTA)}, which has low storage overhead, reduces the number of activation values that must be computed at query time, and guarantees the correctness of the query results. 
NTA builds on the classic threshold algorithm \minor{(CTA)}~\cite{fagin2003optimal}, which supports top-$k$ queries that target arbitrary neuron groups. 
\revision{
However, CTA requires the computation of the activations of all inputs in the dataset at query time.
Because the time to compute the activations by performing DNN inference (not the calculation of the top-$k$ result) dominates the end-to-end query time, CTA would not accelerate our target queries.} 
We argue that for any algorithm to improve query time, it must reduce the number of inputs on which DNN inference is run at query time. DeepEverest achieves this reduction of DNN inference at query time while keeping the storage overhead low by building \minor{NPI} and \minor{using it in \minor{NTA}.} 
Rather than store the raw activations for all neurons, \minor{NPI} partitions the inputs and stores a small amount of information per-partition that is useful when deciding which activation values to recompute at query time. 
\minor{NTA} then uses insights from \minor{CTA} to decide when to terminate as it incrementally \minor{computes} activation values using DNN inference only for small subsets of inputs as needed to answer the query. 
We analyze NTA and show that it is instance optimal for finding the query results of our target queries. 

In addition to its fundamental approach, DeepEverest also includes several important optimizations: 
(1) incremental indexing to avoid large \minor{preprocessing} overhead; 
(2) Maximum Activation Index \minor{(MAI)} to accelerate \textit{top-$k$ most-similar} queries that target maximally activated neurons and \textit{top-$k$ highest} queries; 
(3) \revision{automatic configuration selection}; 
and (4) Inter-Query Acceleration \minor{(IQA)}, which further speeds up sequences of related queries.

In summary, the contributions of this paper are:
\begin{itemize}[itemsep=0pt, topsep=0pt, leftmargin=9pt]
	\item We propose, design, and implement a system called DeepEverest that includes an efficient index structure and an instance optimal query execution algorithm that accelerates \textit{interpretation by example} queries for DNN interpretation while keeping the storage overhead low (\revision{\cref{subsec:overview,,subsec:neuParIdx,,subsec:neuralThresholdAlgorithm,,subsec:proof,,subsec:incrementalIdx}}).
	\item We develop additional optimizations for DeepEverest that further accelerate individual queries (\cref{subsubsec:maxActIdx}), \revision{automatically select a good configuration for the system} (\cref{subsubsec:configSelection}), and accelerate sequences of related queries (\cref{subsubsec:interQueryAcceleration}).
	\item We evaluate DeepEverest on benchmark datasets and models (\cref{sec:evaluation}). We demonstrate that DeepEverest, using less than $20\%$ of the storage of full materialization, significantly accelerates individual \textit{interpretation by example} queries by up to \revision{$63.5{\times}$} and consistently outperforms other methods on multi-query workloads that simulate DNN interpretation processes.
\end{itemize}

\section{Preliminaries} \label{sec:problem}

A DNN consists of layers composed of units, called neurons, connected by edges with associated weights.
Inputs to a DNN are propagated through the layers. 
The output of a neuron is a linear combination of its inputs and their associated edge weights that is optionally transformed by a nonlinear activation function. 
The output of each neuron for a given input is called its \textit{activation value} or \textit{activation}. DNN
interpretation often involves the study of these activations. 
Typical questions that researchers may ask include 
the \textit{interpretation by example} queries mentioned in \cref{sec:introduction}. 
These queries enable researchers to reason about what the DNN learns and identify
how groups of neurons match human-interpretable concepts. 
In this paper, we address the problem of enabling fast queries over activations in a DNN. 
Conceptually, a DNN and an input dataset can be described by the relations \small{\texttt{Neuron(neuronID, layerID, $\ldots$)}} \normalsize and \small{\texttt{Artifact(inputID, neuronID, activation)}}. \normalsize

DeepEverest supports two fundamental classes of queries over activations: \textit{top-$k$ highest} queries that find the top-$k$ inputs that produce the highest activations for a user-specified group of neurons and \textit{top-$k$ most-similar} queries that find the top-$k$ inputs that are most similar to a user-specified target input based on the activations of a user-selected group of neurons.
The rank of an input is decided by a user-specified distance function
(or \minor{a default function}), $\textsc{dist}$. Based on the
user-selected \minor{neuron group}, for \textit{top-$k$ highest}
queries, $\textsc{dist}$ takes as input a set of activations and
measures their magnitude. For \textit{top-$k$ most-similar} queries, $\textsc{dist}$ measures the distance between the input and the target input, and it takes as input a set of absolute differences between the input's activations and the target input's activations. 
Note that \textit{top-$k$ highest} queries can be considered as \textit{top-$k$ most-similar} queries with a hypothetical target input whose activations are infinite for all neurons. 
This distance function $\textsc{dist}$ must be monotonic, i.e., $\textsc{dist}(x_1, x_2, \dots, x_n) \le \textsc{dist}(x_1^\prime, x_2^\prime, \dots, x_n^\prime)$ whenever $x_i \le x_i^\prime$ for each $i$. 
\minor{Monotonicity} is satisfied by common distance functions, such as $l_p$-distances, cosine distance (once transformed to normalized $l_2$-distance), and weighted distances like Mahalanobis distance, among others. 
The \minor{default} in DeepEverest is $l_2$-distance.

\section{Related Work} \label{sec:relatedWork}

\noindent \textbf{DNN Interpretation.} Many approaches have been proposed to interpret the internals of DNNs~\cite{bau2017network, olah2018building, zhou2018revisiting, gonzalez2018semantic, zhou2018interpreting, bau2020understanding, goodfellow2009measuring, zeiler2014visualizing, szegedy2014intriguing, girshick2014rich, zhou2014object, yosinski2015understanding, amershi2015modeltracker, ma2016going, kahng2017cti, caruana1999case, karpathy2015visualizing, nguyen2016synthesizing, papernot2018deep, wallace2018interpreting}. These approaches ask \textit{interpretation by example} queries that return the most similar inputs with respect to the activations of a group of neurons of a given input, or inputs that maximally activate a group of neurons. They motivate the design of DeepEverest, which does not invent new interpretation methods but rather builds novel indexes and algorithms that accelerate the execution of these commonly asked queries. %

\noindent \textbf{Systems for Machine Learning.} Many systems have been proposed to support efficient ML~\cite{vartak2016modeldb, miao2017modelhub, xin2018helix, sparks2017keystoneml}. 
DeepEverest falls into the group of systems that support efficient model diagnosis and interpretation. 
A number of systems like ModelTracker~\cite{amershi2015modeltracker}, CNNVis~\cite{liu2016towards}, and others~\cite{callahan2006vistrails, kahng2016visual, krause2016interacting, kahng2017cti, ludascher2006scientific, kulesza2015principles, yosinski2015understanding} support visual inspection of ML models and features. 
These systems could utilize DeepEverest to accelerate some of the queries used to build the visualizations. 
DeepBase~\cite{sellam2019deepbase} 
lets users identify neurons that have statistical dependencies with user-specified hypotheses. However, it does not support \textit{interpretation by example} queries. 
\minor{MISTIQUE~\cite{vartak2018mistique} accelerates the examination of neuron activations by using storage techniques such as quantization to reduce the storage overhead while sacrificing query accuracy.} 
\minor{These techniques are orthogonal to DeepEverest's and could be incorporated in DeepEverest to further reduce the storage overhead.} 
MISTIQUE also proposes a cost model that captures the trade-off between materialization and recomputation of the activations for different layers and makes materialization decisions accordingly. 
\minor{None of these systems have addressed the problem of accelerating \textit{interpretation by example} queries well because none of them reduce the number of activation values computed during query execution as DeepEverest does.}

\noindent \textbf{Nearest Neighbor Search.} Our target query is a k-nearest neighbor (KNN) search. While many methods exist for exact~\cite{bentley1975multidimensional, guttman1984r, omohundro1989five, liu2006new} and approximate~\cite{datar2004locality, andoni2006near, weiss2008spectral, jegou2010product, johnson2019billion} KNN, the challenge in this paper is different. These KNN methods must know what dimensions will be queried before constructing data structures in that space. In our problem, the dimensions of the KNN search are defined by the neuron group specified only at query time.

\noindent \textbf{Top-K Query Processing.} Top-$k$ query processing is formalized by the seminal work on the threshold algorithm~\cite{fagin2003optimal}. The algorithm scans multiple sorted lists and maintains an upper bound for the aggregate score of unseen objects. Each newly seen object is accessed (by random access) in every other list, and the aggregate score is computed by applying the scoring function to the object’s value in every list. The algorithm terminates after $k$ objects are seen with scores greater than or equal to the upper bound.
Many follow-up approaches propose approximation, optimizations, and extensions~\cite{theobald2004top, ilyas2002joining, bast2006io, akbarinia2007best, pang2010efficient, han2015efficient, zhang2016listmerge}, but they assume that accesses are available to the underlying data sources. 
This assumption does not hold in our problem. 
Storing (on disk) or computing (at query time) the activations required for the top-$k$ queries incurs prohibitive storage overhead or computation overhead respectively.

\section{DeepEverest} \label{sec:coreAlgorithm}

In this section, we first consider baseline approaches, then describe how DeepEverest improves upon these baselines to accelerate query execution while keeping the storage overhead low.

\subsection{Baselines} \label{subsec:baselines}

We first discuss baseline approaches and explain why applying CTA or any KNN algorithm would not improve the query time.

\vspace{0.115em}
\noindent \textbf{\textit{PreprocessAll.}}
The first baseline, \textit{PreprocessAll}, has a high storage cost.
It performs DNN inference on the entire dataset and stores all the activations for all neurons ahead of time.
It executes queries by loading the previously stored activations of the neuron group for all inputs from disk and maintaining a top-$k$ result set.

\vspace{0.115em}
\noindent \textbf{\textit{ReprocessAll.}}
The second baseline, \textit{ReprocessAll}, has a high computation cost.
It has no storage overhead and performs no preprocessing.
It executes queries by computing the activations of the layer being queried by DNN inference on all inputs and maintaining a top-$k$ result set as it performs the \minor{computation}.

\vspace{0.115em}
\noindent \textbf{\textit{LRU Cache.}}
The third baseline, \textit{LRU Cache}, is a disk cache that has a fixed storage budget with a least-recently-used (LRU) replacement policy. This strategy strikes a balance between the storage overhead of \textit{PreprocessAll} and the computation overhead of \textit{ReprocessAll}.
\textit{LRU Cache} maintains a fixed-sized disk cache that stores the activations for queried layers.
A query is executed as in \textit{PreprocessAll} if the activations of the queried layer are present in the disk cache.
Otherwise, it is executed as in \textit{ReprocessAll}. After that, the activations of the queried layer are persisted to the disk cache.
When the size of the disk cache exceeds the storage budget, the cache evicts the activations of the least recently used layer.

\vspace{0.115em}
\noindent \textbf{\textit{Priority Cache.}}
The final baseline, \textit{Priority Cache}, is a technique adapted from MISTIQUE~\cite{vartak2018mistique}. 
It has a fixed-sized disk cache to store the activations for some layers. As a preprocessing step, it uses the storage cost model from \cite{vartak2018mistique} to pick which layers to store, assuming that each layer will be queried the same number of times. Under the storage budget, this cost model prioritizes the layers that save the most query time per GB of data stored. It performs DNN inference on every input and stores the activations for the layers selected ahead of time. A query is executed as in \textit{PreprocessAll} if the activations of the queried layer are present in the disk cache. Otherwise, the query is executed as in \textit{ReprocessAll}.

\vspace{0.115em}
\minor{CTA} could be applied to each baseline by first using the materialized or recomputed activations to construct the \texttt{Artifact} table defined in \cref{sec:problem}.
\texttt{Artifact} is then used to construct a relation in which each row represents an input, and each column represents a neuron and contains the absolute difference between the activation of that row's input and the target input's activation on the column's neuron.
\minor{CTA} can run after sorting the absolute differences \minor{in each column} in ascending order, using any monotonic norm of the absolute differences as the aggregation function.

\revision{
    \dnnBottleneck
}

However, applying \minor{CTA} (or any KNN algorithm) on top of each of the baselines would not improve the query time.
Using it along with \textit{ReprocessAll} would not help because \textit{ReprocessAll} requires running DNN inference on the entire dataset to compute \texttt{Artifact} at query time.
Similarly, applying it on \textit{PreprocessAll} would not improve the query time because generating the relation of absolute differences requires a full scan over \texttt{Artifact} before \minor{CTA} can be applied.
The top-$k$ result could already be computed during the full scan.
\revision{Further, applying it on \textit{PreprocessAll} incurs this strategy's prohibitive storage overhead.}
Applying it on top of \textit{LRU Cache} and \textit{Priority Cache} would not improve the query time for the same reasons as \textit{PreprocessAll} (for layers in the cache) and \textit{ReprocessAll} (for layers not in the cache).
\revision{
Given the prohibitive storage overhead of \textit{PreprocessAll}, any existing method that supports queries for arbitrary neuron groups must compute the activations of the neuron group for all inputs at query time.
\cref{tab:dnnBottleneck} shows the query time breakdown for various baselines. 
The total query time consists of the time for DNN inference to compute the activations of the queried neuron group, the time for building the data structure required for each method (it cannot be computed ahead of time because the neuron group for a query can be arbitrary), and the time to obtain the top-$k$ result.
As the results show, DNN inference is the bottleneck of query execution.
Therefore, any method that does not reduce the number of inputs fed into the DNN at query time will perform similarly to \textit{ReprocessAll}. 
Hence, the query time of \textit{ReprocessAll} can represent that of these more advanced methods.
}

\subsection{Overview of DeepEverest} \label{subsec:overview}

As described above, directly applying \minor{CTA} does not improve the query time because \texttt{Artifact} must be fully computed \revision{for the neuron group} at query time, which requires DNN inference on all inputs. Query execution can be significantly accelerated by avoiding running the DNN on inputs that will not be one of the top-$k$ results.

We design and build a novel index, called the Neural Partition Index \minor{(NPI)} (\cref{subsec:neuParIdx}), and a query execution algorithm, called the Neural Threshold Algorithm \minor{(NTA)} \revision{(\cref{subsec:neuralThresholdAlgorithm})}. 
\minor{NTA} is a modified threshold algorithm. 
In contrast to \minor{CTA}, NTA does not require all activations of all inputs before it starts. 
It utilizes \minor{NPI} to progressively access and perform DNN inference on only the inputs that are possibly in the top-$k$ results. 
NTA overcomes the bottleneck of query execution, DNN inference, by reducing the number of inputs on which DNN inference is performed at query time, while guaranteeing the correctness of the top-$k$ results and introducing only tolerable storage overhead. 
Moreover, we show the instance optimality of NTA (\cref{subsec:proof}) and propose various additional optimizations (\cref{subsec:incrementalIdx,,subsec:optimizations}).

\subsection{Neural Partition Index \minor{(NPI)}} \label{subsec:neuParIdx}

\revision{A key goal of DeepEverest is to support queries that target arbitrary neuron groups. Existing partitioning or indexing methods (e.g., \textit{K-D Tree, locality-sensitive hashing}) must know which neuron group will be queried ahead of time and construct data structures in that space. In contrast, DeepEverest constructs indexes for each neuron separately and therefore is able to answer queries for arbitrary neuron groups.}
The activations in a DNN can conceptually be represented by the \texttt{Artifact} relation introduced in \cref{sec:problem}.  
\revision{
Recall that \texttt{neuronID} is the identifier for a neuron in the DNN and \texttt{inputID} is the identifier for an input in the dataset.} 
DeepEverest \minor{conceptually} builds an index on the search key \texttt{(neuronID, activation)} and supports queries that return the \texttt{inputID}s for a given \texttt{neuronID} and range of \texttt{activation} values.
\revision{To avoid materializing activations, DeepEverest builds an index on \texttt{(neuronID, PID)} instead,} \revision{where \texttt{PID} (\texttt{partitionID}) is the identifier of a range-partition over activation values for a neuron.}
\revision{
DeepEverest builds equi-depth partitions instead of equi-width partitions because the activation values are usually highly skewed, and equi-depth partitions adapt better to skewed distributions.
Partition $0$ contains the largest activations.} 
The index then supports efficient lookups for a given \minor{\texttt{(neuronID, PID)}} combination. 
The index returns the set of \texttt{inputID}s whose activations for the given \texttt{neuronID} belong to the partition identified with \minor{\texttt{PID}}.
Moreover, the index also supports queries that return the \minor{\texttt{PID}} for a given \texttt{(neuronID, inputID)} combination.
We call this structure the Neural Partition Index \minor{(NPI)} and denote the queries with \revision{\texttt{getInputIDs(neuronID, PID)} and \texttt{getPID(neuronID, inputID)}.}
Additionally, in NPI, for each partition, DeepEverest stores the \textit{lower bound} \revision{and \textit{upper bound}} of the activations in that partition and supports queries that ask for them.
We denote such queries with \minor{\texttt{lBnd(neuronID, PID)}} and \revision{\texttt{uBnd(neuronID, PID)}}.
The number of partitions, $nPartitions$, is configurable and discussed further in \ref{subsubsec:configSelection}.

\exampleNeuralPartitionIndex

There are two approaches to implementing the index. The first approach would be to maintain a set of buckets, each identified with a unique \minor{\texttt{(neuronID, PID)}} combination as the key, and, for each bucket, maintain a list of \texttt{inputID}s. The second approach, which DeepEverest uses, is to maintain a list of \texttt{(neuronID, inputID)} pairs as keys, and, for each entry, store the \texttt{PID}. 
\texttt{neuronID}, \texttt{inputID}, \minor{and \texttt{PID}} are integers, so rather than building a B-tree or a hash index over the keys,
we create an optimized index structure using an array where \texttt{neuronID} and \texttt{inputID} act as offsets for lookups in the array. 
This enables DeepEverest to only store the values and therefore avoid the cost of storing the keys. 
\cref{fig:exampleNeuralPartitionIndex} illustrates \minor{NPI} for an example dataset with three partitions.

During preprocessing, DeepEverest runs DNN inference on all inputs once to build \minor{NPI} for every neuron, \revision{which requires sorting.} 
The time complexity to compute NPI once the activations are computed is $O(nNeurons \cdot nInputs \cdot log_2(nInputs))$, but the main source of overhead is DNN inference, which is a cost proportional to the number of inputs. 
The method that DeepEverest adopts is more space-efficient than building an index over \minor{\texttt{(neuronID, PID)}} pairs because it costs $nNeurons \cdot nInputs \cdot \log_{2}(nPartitions)$ bits rather than $nNeurons \cdot nInputs \cdot \log_{2}(nInputs)$ bits, where $nPartitions << nInputs$. 
\minor{NPI} also has much smaller storage overhead compared to fully materializing all activation values. 
A \minor{\texttt{PID}} takes less storage than an activation value because a \minor{\texttt{PID}} only costs $log_2(nPartitions)$ bits, while an activation value is usually a 32-bit floating point.
For example, if DeepEverest has $8$ partitions for each neuron, representing a \minor{\texttt{PID}} costs 3 bits, which is less than $10\%$ of the storage cost of full materialization.
\revision{Storing the lower and upper bounds costs $nNeurons \cdot nPartitions \cdot 2 \cdot 32$ bits, which is normally negligible compared to the cost of storing the \minor{\texttt{PID}s}.}

\vspace{-0.4em}
\subsection{Neural Threshold Algorithm \minor{(NTA)}} \label{subsec:neuralThresholdAlgorithm}

\exampleDParOrd
\exampleToRunDistTop

\textbf{Notation.} We denote with $N$ the set of all neurons in the DNN and with $D$ the input dataset. 
A neuron is denoted with $n {\in} N$ and an input with $x {\in} D$. 
\revision{$x$ is the \texttt{inputID}.} 
The user issues a query: \minor{\texttt{topk(}$s, G, k,\textsc{dist}$\texttt{)}}, where $s {\in} D$ is the sample input (also known as the target input) of interest to the user. 
\minor{$G {\subseteq} N$} is a set of neurons \minor{from a single layer in $N$.} $k$ is the desired number of query results, and \textsc{dist} is the distance function. 
This function computes the distance between the set of activations of $s$ and $x$ looking only at the neurons in \minor{$G$}. 
\cref{tab:notation} lists our frequently used notation.

\notationTable

\minor{NTA} returns the set of top-$k$ inputs that are closest to the target input when considering only the neurons in \minor{$G$}. This set of top-$k$ inputs can be defined as a set $top \subseteq D$ of $k$ inputs. $top$ is initially empty and is conceptually built incrementally by identifying and adding to $top$ the next input that satisfies:
\minor{
\useshortskip
\begin{equation}
  \arg\min_{x {\in} D \setminus top} \left\{ dist(s, x, G) \right\}
\end{equation}
}
\noindent We further denote with \minor{$g_i$} (or $i$ when clear) the $i$-th neuron in set \minor{$G$}, and with \texttt{act}$(i, x)$ the activation of neuron $g_i$ on input $x$. 
\revision{$g_i$ is the \texttt{neuronID}.} 
For each neuron $n {\in} N$, \minor{NPI} includes the set of partitions, \minor{$P_n$}.
We denote a single partition for neuron $n$ with \minor{$p {\in} P_n$}, \revision{and $p$ is the \texttt{PID}.}
\revision{We denote the lower and upper bounds of this partition $p {\in} P_n$ with \texttt{lBnd}$(n, p)$ and \texttt{uBnd}$(n, p)$.} 

\vspace{0.115em}
\minor{NTA} proceeds as follows,

\vspace{0.115em}
\noindent \textbf{Step 1: Load indexes.} We assume that \minor{NPI} is initially on disk. This step reads the \minor{NPI} for the neurons \minor{$g_i {\in} G$} from disk. The index holds the set of partitions, their \revision{lower and upper bounds}, and the \minor{\texttt{PID}s} of each input for each \minor{$g_i {\in} G$}. %

\vspace{0.115em}
\noindent \textbf{Step 2: Compute target activations.}
For each \minor{$g_i {\in} G$}, compute \texttt{act}$(i, s)$, the activation value for input $s$ and neuron \minor{$g_i$} by running DNN inference on input $s$.
A single inference pass is sufficient to compute the activations for all neurons in \minor{$G$}.
    
\vspace{0.115em}
\noindent \textbf{Step 3: Order partitions.}
This step computes the order by which the partitions are accessed by NTA for each neuron.
\revision{Let $\texttt{sPID}(i)$ denote the \texttt{PID} to which $s$ belongs for neuron $g_i$.}
For each neuron \minor{$g_i {\in} G$} and partition $p {\in} P_i$, compute $dPar(i, p)$ as,
\revision{
\useshortskip
\begin{equation} \label{eq:dpar}
    dPar(i, p) = \begin{cases}
                    0, & p = \texttt{sPID}(i) \\
                    \texttt{lBnd}(i, p) - \texttt{act}(i, s), & p < \texttt{sPID}(i) \\
                    \texttt{act}(i, s) - \texttt{uBnd}(i, p), & p > \texttt{sPID}(i)
                \end{cases}
\end{equation}
}
\noindent which is the distance between the target input's activation value for neuron $g_i$ and the \revision{closest} activation value in partition $p$. For each neuron $g_i$, sort the partitions \minor{in $P_i$} on their $dPar(i, p)$ values \revision{in ascending order} and put them in a list, denoted with $ord(i)$.
\revision{Later steps will process the partitions in the order specified by $ord(i)$.}

\textit{
Example: To illustrate the first three steps, consider a query \texttt{topk(}$x5, \{R1, R2, R3\}, 2, l1$-\texttt{distance)} that finds the top-$2$ most similar inputs to $x5$ based on the activations of $\{R1, R2, R3\}$, using the example dataset in \cref{fig:exampleNeuralPartitionIndex}.
In this example, \minor{$s {=} x5$, $G {=} \{R1, R2, R3\}$}.
Step 1 reads from disk \minor{PID, lBnd}, \revision{and uBnd} shown in \cref{fig:exampleNeuralPartitionIndex}.
Step 2 runs DNN inference to compute the activations for $x5$, $(\texttt{act}(R1, x5), \texttt{act}(R2, x5), \texttt{act}(R3, x5)) {=} (1.1, 1.1, 1.2)$. 
Step 3 computes $dPar$ and $ord$ for $\left\{ R1, R2, R3 \right\}$, as shown in \cref{fig:exampleDParOrd}.
}

\vspace{0.115em}
\noindent \textbf{Step 4: Find top-k.}
This step runs the modified threshold algorithm. It starts with the partitions to which the target input belongs, and it expands its search from there. Unlike \minor{CTA}, this step incrementally computes the activations for candidate inputs and does so in batches to get good GPU performance. This step proceeds as follows, 

Starting with an index $c = 0$:

\vspace{0.115em}
\noindent \textbf{Step 4 (a):}
            For each neuron \minor{$g_i$}, maintain a set $toRun_{i}$ that contains the inputs whose activations should be computed.
            Access $ord(i, c)$ to get the partition that contains the next most similar inputs.
            Query \minor{NPI} to get the \texttt{inputID}s that belong to this partition $ord(i, c)$ and add them to $toRun_{i}$, i.e., \minor{$toRun_{i} {\gets} \texttt{getInputIDs}(i, ord(i, c))$}.

            \textit{Example: as shown in \cref{subfig:toRun}, when $c = 0$, $toRun_{R1, 0} = \left\{ x4, x5 \right\}$ because $ord(R1, 0) = 2$.}

\vspace{0.115em}
\noindent \textbf{Step 4 (b):}
            For each neuron $g_i$, compute the activations for the inputs in $toRun_{i}$ (excluding those that have already been computed) by running DNN inference in batches. $toRun_{i}$ is cleared after DNN inference.
            Note that this inference step computes the activations for \textit{all} neurons in the neuron group being queried.
            Compute the distance between each newly computed input $x$ and the target input $s$ as \minor{$dist(s, x, G) = \textsc{dist}(|\texttt{act}(0, x) - \texttt{act}(0, s)|, \dots, |\texttt{act}(|G| - 1, x) - \texttt{act}(|G| - 1, s)|)$}.
            Update $top$ if \minor{$dist(s, x, G)$} is one of the $k$-smallest NTA has seen so far, i.e., input $x$ is one of the $k$-most similar inputs to the target input $s$ seen so far.
            Ties are broken arbitrarily.

            \textit{Example: as shown in \cref{subfig:dist}, when $c{=}0$, the activations of inputs $x2, x4$ are computed ($x5$ was computed in Step 1). The distances from $x2$ and $x4$ to $x5$ are $1.5$ and $0.3$, respectively.}

\vspace{0.115em}
\noindent \textbf{Step 4 (c):}
            Maintain a range of seen activations for inputs from $toRun_i$ for each neuron $g_i$, which is the range of activations such that NTA has seen every input with an activation in the open interval of this range. It is possible that NTA has seen one or more inputs from other neurons' $toRun$ sets with activations outside of this range. However, the open interval of this range denoted by $(minBoundary_i, maxBoundary_i)$ only contains the \minor{activations for the inputs that NTA is guaranteed to have seen.}

            Let $minDist_i$ be the shorter distance from the boundaries of this range to the target input for each neuron $g_i$: \minor{$minDist_i = \min \left\{ F_{i} {\cdot} |minBoundary_{i} {-} \texttt{act}(i, s)|, V_{i} {\cdot} |maxBoundary_{i} {-} \texttt{act}(i, s)| \right\}$}, where $F_{i}$ is an indicator function that indicates whether NTA has seen the last partition (inputs with lowest activations) of neuron $g_i$, and \minor{$V_{i}$} is another indicator function that indicates whether the $0$-th partition (inputs with highest activations) of neuron $g_i$ has been seen.
            Specifically, $F_{i} {=} \infty$ when the last partition of neuron $g_i$ has been seen; $F_{i} {=} 1$ otherwise. \minor{$V_{i} {=} \infty$} when the first partition of neuron $g_i$ has been seen; \minor{$V_{i} {=} 1$} otherwise.
            Define the threshold to be,
            \minor{
            \useshortskip
            \begin{equation} \label{eq:threshold}
                t = \textsc{dist}(minDist_{0}, minDist_{1}, \dots, minDist_{|G| - 1})
            \end{equation}
            }
            \noindent The threshold, $t$, represents the smallest possible distance to $s$ from any unseen input.
            Hence, the termination condition is,
            \minor{
            \useshortskip
            \begin{equation} \label{eq:termination}
                \max_{x \in top} \left\{ dist(s, x, G) \right\} \leq t
            \end{equation}
            }
            \noindent where $\max_{x \in top} \left\{ dist(s, x, G) \right\}$ represents the maximum distance to the target input $s$ in the current top-$k$ result set. As soon as this inequality holds, halt and return $top$ as the query results.

            \textit{Example: as shown in \cref{subfig:termination}, $minBoundary_{i}$, $maxBoundary_{i}$ and $minDist_{i}$ are maintained and calculated for $\left\{ R1, R2, R3 \right\}$. For example, when $c = 0$, $minBoundary_{R1} = 1.1$, $maxBoundary_{R1} = 1.2$. Since NTA has seen the last partition (2) and has not seen the first partition (0), $F_{R1} = \infty, V_{R1} = 1$. Therefore, $minDist_{R1} = |maxBoundary_{R1} - act_{R1, x5}| = |1.2 {-} 1.1| = 0.1$.}

            \textit{When $c = 0$, \minor{${t = 0.2 < 1.5 = \max_{x \in top} \left\{ dist(s, x, G) \right\}}$}, so NTA does not halt.
            When $c = 1$, \minor{$t = 1.7 \ge 1.5 = \max_{x \in top} \left\{ dist(s, x, G) \right\}$}, so NTA halts and returns $top$ as the query result. It is worth noting that the cost of DNN inference on $x0$ is not incurred because it is impossible for $x0$ to be one of the top-$2$ results.}

\vspace{0.115em}
\noindent \textbf{Step 4 (d):}
            Increment $c$ by $1$. Repeat Step 4 (a) - (d) until all partitions have been seen or the halting condition in Step 4 (c) is satisfied.

\neuralThresholdAlgorithm
The pseudocode is shown in \cref{alg:neural-threshold-algorithm}.

The key innovation of \minor{NTA} compared to \minor{CTA} is in its processing of the inputs partition-by-partition using \minor{NPI} until the termination condition is met.
\minor{NTA} runs DNN inference on only the necessary partitions of inputs for it to be certain that it has the precise top-$k$ results when it terminates.
This approach significantly reduces the number of inputs on which DNN inference is performed at query time compared to computing the activation values for all inputs at query time.
\minor{It further improves query performance by utilizing batch processing on GPUs \cite{nvbatch}. 
Inputs that share a partition are sent to the DNN for inference all at once.}
\minor{NTA} along with NPI also has much smaller storage overhead compared to fully materializing the activations for all inputs.

\revision{
\vspace{-0.4em}
\subsection{Instance Optimality of NTA} \label{subsec:proof}

In this section, we investigate the instance optimality of NTA, which corresponds to the optimality in every instance, rather than just the worst or average case. Fagin's paper~\cite{fagin2003optimal} shows that CTA is instance optimal for finding the top-$k$ items over all algorithms (excluding those that make very lucky guesses) and over all legal databases. We build on that proof to show the same for NTA.

Following the definitions in~\cite{fagin2003optimal}, let $\mathbb{A}$ be a set of algorithms that answer our target queries, and $\mathbb{D}$ be a set of combinations of datasets and DNN models that are legal inputs. 
Let $cost(\mathcal{A}, \mathcal{D})$ be the number of inputs in the dataset of $\mathcal{D}$ accessed by $\mathcal{A} \in \mathbb{A}$ when running $\mathcal{A}$ on $\mathcal{D} \in \mathbb{D}$.
An algorithm $\mathcal{B}$ is instance optimal over $\mathbb{A}$ and $\mathbb{D}$ if $\mathcal{B} \in \mathbb{A}$ and if for every $\mathcal{A} \in \mathbb{A}$ and every $\mathcal{D} \in \mathbb{D}$,
\useshortskip
\begin{equation}
    cost(\mathcal{B}, \mathcal{D}) = O(cost(\mathcal{A}, \mathcal{D}))
\end{equation}
We show the following theorem.
\useshortskip
\vspace{-0.25em}
\begin{theorem}
    NTA is instance optimal for finding the top-$k$ inputs over all algorithms (excluding those that make very lucky guesses) and over all legal combinations of datasets and DNN models.
\end{theorem}
\vspace{-0.25em}
The proof follows similarly to that of Theorem 6.1 in~\cite{fagin2003optimal}. We bound the maximal number of inputs accessed by NTA with an additive constant over CTA's maximal sequential access depth.
\vspace{-0.35em}
\begin{proof}
    For any $\mathcal{D} {\in} \mathbb{D}$, assume that we already have the relation of sorted absolute differences (denoted with \texttt{AbsDiff}) between the activations of all inputs and the activation of the target input.
    Assume that CTA halts at depth $d_i$ for each neuron $g_i$ when running on \texttt{AbsDiff}, i.e., $d_i$ is the number of inputs seen via sequential accesses for neuron $g_i$. 
    As in~\cite{fagin2003optimal}, it suffices to bound the maximal number of inputs accessed by NTA for any $g_i {\in} G$.
    Let $d = \max_i d_i$.
    For the duration of this proof, let $x_j$ denote the input at depth $j$ in \texttt{AbsDiff} for a neuron $g_i$ that reaches depth $d$ when running CTA.
    Let $R$ be the partition size in NPI.
    We will show that NTA will have accessed $x_j (\forall 1 {\le} j {\le} d)$ and terminate in at most $d + 2 R$ accesses.

    Recall that we denote a single partition for neuron $g_i$ with $p {\in} P_i$. $p$ is the partition identifier, \texttt{PID}, and $\texttt{sPID}(i)$ is the \texttt{PID} of the target input $s$ for neuron $g_i$.
    In NPI, for $g_i {\in} G$, we say that a partition $p {\in} P_i$ is an \textit{above} partition if $p < \texttt{sPID}(i)$; a partition $p {\in} P_i$ is an \textit{under} partition if $p \ge \texttt{sPID}(i)$.
    We say that an input is an \textit{above} input if it belongs to an \textit{above} partition; an input is an \textit{under} input if it belongs to an \textit{under} partition.
    As in \cref{eq:dpar}, an \textit{above} partition uses the input whose activation is the lower bound of the partition as its representative; an \textit{under} partition uses the input whose activation is the upper bound as its representative (the only exception is that the partition $p = \texttt{sPID}(i)$ uses the target input, $s$, as its representative).

    Without loss of generality, assume that $x_d$ is an \textit{above} input. Let $a$ be the greatest $j {<} d$ where $x_j$ is an \textit{under} input.
    Let $b$ be the least $j {>} d$ where $x_j$ is an \textit{under} input.
    When NTA has accessed $x_j (\forall 1 {\le} j {\le} d)$ and confirmed that $x_b$ is more distant from $s$ than $x_d$, it knows that it has the correct top-$k$ results and can then terminate. This is equivalent to the termination condition in \cref{subsec:neuralThresholdAlgorithm}.

    \textbf{Case 1}: When $x_a$ is accessed by NTA after $x_d$, we will show at most $R$ \textit{above} inputs, $x_j (j {>} d)$, are accessed before $x_a$.
    Assume towards contradiction that more than $R$ \textit{above} inputs $x_j (j {>} d)$ are accessed before $x_a$.
    There would be a representative of an \textit{above} partition, $x_{d^{\prime}}$ where $d^{\prime} \ge d$. Let $x_{a^{\prime}}$ be the representative of the partition containing $x_{a}$.
    Thus, $a^{\prime} \le a < d \le d^{\prime} \implies a^{\prime} < d^{\prime}$.
    This is a contradiction because the \textit{above} partition with $x_{d^{\prime}}$ as its representative was chosen to be accessed earlier than the \textit{under} partition with $x_{a^{\prime}}$ as its representative, which implies $d^{\prime} \le a^{\prime}$.

    Since \textit{under} partitions are accessed in order w.r.t. $x_{a}$, the number of \textit{under} inputs, $x_{j} (j {>} d)$, accessed by NTA is also at most $R$, i.e., the \textit{under} inputs co-located on the partition containing $x_{a}$.

    Note that if $x_{a}$ and $x_{b}$ belong to the same partition, NTA will terminate after this partition.
    If $x_{a}$ and $x_{b}$ belong to different partitions, then $x_{b}$ is the representative of its partition.
    After confirming that $x_{b}$ is more distant from the target input $s$ than $x_{d}$, NTA knows that it does not need to access any further partitions (including the partition containing $x_{b}$) and terminates.
    In either scenario, the number of \textit{above} $x_{j} (j {>} d)$ accessed by NTA is at most $R$, and the number of \textit{under} $x_{j} (j {>} d)$ accessed is also at most $R$.
    Hence, the total number of accesses made by NTA is at most $d + 2 R$.

    \textbf{Case 2}: When $x_{b}$ is accessed by NTA before $x_{d}$, there can be at most $R$ \textit{under} inputs, $x_{j} (j {>} d)$, that are accessed before $x_{d}$. %
    Furthermore, the number of \textit{above} inputs, $x_{j} (j {>} d)$, accessed is at most $R$, i.e., the \textit{above} inputs co-located on the partition containing $x_{d}$.
    Hence, the total number of accesses made by NTA is at most $d + 2 R$.
    The proof follows similarly to Case 1.

    \textbf{Case 3}: When $x_{a}$, $x_{b}$ are accessed by NTA in order w.r.t. $x_{d}$, $x_{b}$ must be the representative for its partition.
    After NTA processes the partition containing $x_{d}$, it will recognize that $x_{b}$ is more distant than $x_{d}$ and terminate.
    Thus no \textit{under} $x_{j} (j {>} d)$ will be explicitly accessed.
    The \textit{above} partitions are accessed in order w.r.t. $x_{d}$, so the number of \textit{above} inputs, $x_{j} (j {>} d)$, accessed is at most $R$, i.e., the inputs co-located on the partition containing $x_{d}$.
    Hence, the total number of accesses made by NTA is at most $d + R$.

    Since the three cases above exhaust all possibilities, this proves that for each neuron in $G$, NTA will have accessed $x_{j} (\forall 1 {\le} j {\le} d)$ and terminate in at most $d + 2 R$ accesses. \qedhere
\end{proof}

}

\vspace{-0.5cm}
\subsection{Incremental Indexing} \label{subsec:incrementalIdx}

\minor{As shown in \cref{sec:evaluation}}, the DeepEverest approach described so far achieves excellent query execution times with low storage overhead. 
This approach, however, incurs a potentially high preprocessing cost, especially for large datasets and models. 
Before executing any query, DeepEverest needs to compute the activations for all neurons and \revision{all inputs} by running DNN inference. 
It then needs to construct the indexes for all layers and persist them to disk. 

\revision{To address this challenge, we propose building the indexes incrementally as queries execute so that preprocessing is performed \textit{only for the layers users query}.} 
With this approach, DeepEverest performs no preprocessing ahead of time.
When the user submits a query, if the indexes of the queried layer are available on disk, DeepEverest proceeds as described in \cref{subsec:neuralThresholdAlgorithm,,subsubsec:maxActIdx};
otherwise, DeepEverest computes the activations of the queried layer by running DNN inference on all inputs. 
While doing so, it computes the query answer and returns it to the user.
DeepEverest then constructs the indexes for the layer and persists them to disk. 
\revision{
Note that DNN inference is performed starting from the first layer instead of a previously queried layer each time DeepEverest constructs the indexes for a layer because only the indexes of the queried layers are stored on disk. 
With this approach, the costs of index computation and persistence for each layer are incurred once the first time that layer is queried, and only if that layer is queried.
}

In \cref{subsec:evalCumulativePerf}, we show that DeepEverest with this incremental approach significantly outperforms other methods on multi-query workloads. 
While DeepEverest must do extra preprocessing to build and store its indexes compared with caching the activations directly, it accelerates significantly more queries because it is able to store the indexes for significantly more layers given a storage budget.

\vspace{-0.6em}
\subsection{Optimizations} \label{subsec:optimizations}

In this section, we present several important optimizations that further improve the performance of DeepEverest. 
The first optimization, described in \cref{subsubsec:maxActIdx}, accelerates two common types of top-$k$ queries. 
The second optimization, described in \cref{subsubsec:configSelection}, automatically \revision{selects DeepEverest's configuration}. 
The third optimization, described in \cref{subsubsec:interQueryAcceleration}, accelerates sequences of related queries, as may occur during data exploration.

\vspace{-0.5em}
\subsubsection{Maximum Activation Index \minor{(MAI)}} \label{subsubsec:maxActIdx}
For a given target input and a layer in a DNN, the \textit{maximally activated neurons} are those neurons in the layer for which the activation values for the target input are the highest. DNN interpretation often involves examining such maximally activated neurons \cite{srivastava2013compete, zeiler2014visualizing, zhou2014object, bau2017network} because they respond to the input the most and have the greatest impact on the DNN output.
A common set of \textit{top-$k$ most-similar} queries ask to find the top-$k$ similar inputs to a target input based on a neuron group consisting of these maximally activated neurons.

To accelerate these \textit{top-$k$ most-similar} queries that target maximally activated neurons as well as \textit{top-$k$ highest} queries, we introduce a straightforward yet effective optimization. 
The idea is for DeepEverest to store, for each neuron, a fraction of the highest activation values together with the corresponding \texttt{inputID}s. 
We call this data structure the Maximum Activation Index \minor{(MAI)}, and \minor{denote this fraction of \texttt{(activation, inputID)} pairs for a given \texttt{neuronID} with \texttt{MAI(neuronID)}}. 
This fraction automatically becomes each neuron's $0$-th partition. 
We denote the fraction of the inputs stored in MAI with $ratio$, which is a configurable parameter and discussed further in \cref{subsubsec:configSelection}. 

\exampleMaximumActivationIndex

DeepEverest utilizes \minor{MAI} during query execution to further reduce the number of inputs on which DNN inference is performed, if possible. DeepEverest now has more detailed knowledge of which inputs are most similar to the target input, rather than just the high-level knowledge that the inputs are in the same partition. We observe empirically that the activation values of the \textit{maximally activated neurons} for an input are often likely to be in the top activations stored in \minor{MAI}, and thus \minor{MAI} is effective in improving the query time.
DeepEverest modifies query execution described in \cref{subsec:neuralThresholdAlgorithm} of partition $0$ to incorporate this information as follows: 
it first finds the neurons for which the target input is in \minor{MAI}. For these neurons, DeepEverest sorts the other inputs in the $0$-th partition by their distances to the target input $s$. 
Rather than performing DNN inference on all inputs in \minor{MAI} for each neuron, DeepEverest builds a global $toRun$ set by adding the most similar inputs from all of these neurons until the batch size is reached.
Step 4(c) in \cref{subsec:neuralThresholdAlgorithm} is modified to compute $minDist_i$ as $ \min \left\{|minBoundary_{i} {-} act_{i, s}|, H_{i} \cdot |maxBoundary_{i} {-} act_{i, s}| \right\}$, where $H_{i}$ is an indicator function that indicates whether NTA has seen the input with the highest activation in \texttt{MAI(i)}. 
$H_{i} {=} \infty$ when highest activation of neuron $g_i$ has been seen; $H_{i} {=} 1$ otherwise. 
The neurons $g_i$ for which $s$ is not in \texttt{MAI(i)} contribute $0$ to the threshold calculation. 
\cref{fig:exampleMaximumActivationIndex} illustrates an example.

\vspace{-0.5em}
\subsubsection{\revision{Automatic Configuration Selection}} \label{subsubsec:configSelection}
Given a storage budget, DeepEverest must allocate it between \minor{NPI} and \minor{MAI}.
\minor{It uses a heuristic algorithm to achieve this}.
\revision{
A greater $nPartitions$ leads to smaller partitions, which is key to generally better performance (see \cref{subsec:evalConfigSelect}) on queries that target any kind of a neuron group (see \cref{subsec:evaluationSetup}), while a larger $ratio$ accelerates only the two types of queries mentioned in \cref{subsubsec:maxActIdx}.
Hence, DeepEverest first picks a value for $nPartitions$ and then sets the value of $ratio$.
}

Intuitively when partitions are smaller, DNN inference is performed on fewer inputs not in the top-$k$ because DeepEverest processes inputs partition-by-partition. However, if the partitions are smaller than the optimal batch size, DeepEverest will not leverage the full GPU \minor{parallelism}. %

Given a storage budget, $budget$ (in bytes), and a batch size, $batchSize$, DeepEverest sets $nPartitions$ to be the maximum power of two (to utilize all bits) that satisfies $nPartitions \le nInputs / batchSize$ and $cost(nPartitions) < budget$.
$cost(nPartitions)$ is the bytes consumed by storing \minor{NPI} and is calculated as $nNeurons \cdot nInputs \cdot \log_{2}(nPartitions) / 8$.
$batchSize$ is set to the value that achieves the highest throughput for the DNN. 
Given the remaining storage budget, DeepEverest sets $ratio$ to be the maximum value that satisfies $cost(ratio) \le budget - cost(nPartitions)$, where $cost(ratio)$ is the bytes consumed by storing \minor{MAI} and calculated as $ratio \cdot nInputs \cdot nNeurons \cdot 4 \cdot 2$, since \texttt{activation} and \texttt{inputID} are 4 bytes each. When there is no remaining storage budget after selecting $nPartitions$, DeepEverest sets $ratio$ to $0$.

\vspace{-0.5em}
\subsubsection{Inter-Query Acceleration (IQA)} \label{subsubsec:interQueryAcceleration}
Inter-Query Acceleration (IQA) is an optimization technique to accelerate sequences of related queries, as may occur during DNN interpretation. 
As an example, imagine that a user finds a misclassified image. The user may want to first see the maximally activated neurons in a layer for the image and then find images with similar maximally activated neurons. The user may then decide to change how many neurons they are looking at, e.g., go from the top-3 neurons to the top-4 neurons.
Queries that overlap in neurons present an opportunity for further optimization as activation values can be reused for related queries.

With IQA, DeepEverest leverages an in-memory cache that contains recently used activation values to reduce the number of inputs that it must run DNN inference on at query time. 
Note that this in-memory cache is different from the disk caches described in \cref{subsec:baselines}.
During query execution, DeepEverest inserts the activation values of each input processed by \minor{NTA} into the cache. Instead of caching only the activation values for the neuron group being queried, it caches the activations for \textit{all} neurons in the queried layer. 
This enables DeepEverest to utilize the cache for future related queries that target a different group of neurons in the same layer. 
DeepEverest adopts a most recently used (MRU) replacement policy for the in-memory cache. This is because DeepEverest processes partitions in order from most similar to the target input to least similar, and it seeks to prioritize keeping the activations from the most similar partitions in the cache.
We show in \cref{subsec:evalInterQuery} that given a small in-memory cache budget, DeepEverest with \minor{IQA} achieves up to \revision{$8{\times}$} faster query times than DeepEverest without it.

\section{Evaluation} \label{sec:evaluation}

DeepEverest is implemented in Python, using C++ to build the indexes.
We evaluate it against the baselines described in \cref{subsec:baselines}.

\performance

\subsection{Evaluation Setup} \label{subsec:evaluationSetup}

\textbf{Datasets and models.} We evaluate DeepEverest on two sets of well-known datasets and models.
The first, called \cifar, uses as inputs \minor{10,000 images from the test set of CIFAR10 \cite{krizhevsky2009learning}, and uses a VGG16 network~\cite{simonyan2014very, liu2015very}.}
The second, called \imagenet, uses as inputs \minor{10,000 images from the validation set of ImageNet~\cite{russakovsky2015imagenet}, and uses a ResNet50 model~\cite{he2016deep}.}
These two sets of models and datasets complement each other in terms of model and input size and DNN inference cost. 
\minor{In all experiments, we pre-load the entire input dataset into memory.} 
We set $batchSize$ for each model as the value that achieves the highest inference throughput (128 for \cifar; 64 for \imagenet).

\vspace{0.115em}
\noindent \textbf{Query generation.} To generate queries, we consider 3 types of layers: \textit{early}, \textit{mid}, and \textit{late}. 
\minor{For \cifar, these correspond to activation layers 2, 7, and 13.} 
\minor{For \imagenet, we use activation layers 2, 25, and 48.} 
Given an input and a layer, we consider the following types of neuron groups: 
(a) \textit{Top}: the maximally activated neurons for the given input in the layer; 
and (b) \minor{\textit{RandHigh}}: neurons randomly picked from the top half of non-zero neurons for the given input. 
We further consider \textit{small}, \textit{medium}, and \textit{large} neuron groups consisting of 1, 3, and 10 neurons.
Finally, based on the neuron groups, we use the following query types: 
(a) \minor{\textit{FireMax}: \textit{top-$k$ highest} query; }
(b) \minor{\textit{SimTop}: \textit{top-$k$ most-similar} query based on a \textit{Top} neuron group; }
and (c) \minor{\textit{SimHigh}: \textit{top-$k$ most-similar} query based on a \textit{RandHigh} neuron group. }
We randomly select inputs from each dataset to generate \minor{\textit{SimTop}} and \minor{\textit{SimHigh}} queries. 

In all experiments, we set $k {=} 20$, which is a reasonable number of results for a user to inspect for a query. 
With a smaller $k$, we expect DeepEverest to achieve larger speedups because it will process fewer inputs and therefore return the results faster, while the query times of baselines will remain similar since they still need to recompute or load all the activations and maintain the query results. 
With a larger $k$, the overall speedups could degrade, but DeepEverest can incrementally return the top-$k$ query results, as discussed in \cref{sec:discussion}. 
Therefore, the perceived query time is still significantly improved. 
We use $l_2$-distance as the distance function. 
All numbers reported are median values of five queries on random inputs for each query configuration (e.g., query type: \minor{\textit{FireMax}}, neuron group size: 3, layer: \textit{late}). 

\vspace{0.115em}
\noindent \textbf{Machine configuration.} All experiments are run on an AWS EC2 p2.xlarge instance, which has an Intel Xeon E5-2686 v4 CPU running at 2.3 GHz, with 61 GB of RAM, an NVIDIA K80 GPU with 12 GB of GPU memory, \minor{and EBS gp3 volumes for disk storage.}

\subsection{Fundamental Space-Time Tradeoff} \label{subsec:evalEndToEnd}
\CTTIncrementalIdx

We first evaluate the fundamental trade-off that DeepEverest achieves in terms of storage space and query execution time for individual queries. 
In this experiment, the only optimization DeepEverest uses is \minor{MAI} described in \cref{subsubsec:maxActIdx}. 
We first precompute and store the indexes for all layers before executing the benchmark queries.
DeepEverest has a storage budget of $20\%$ of \textit{PreprocessAll}, and selects $nPartitions$ and $ratio$ using the algorithm described in \cref{subsubsec:configSelection}. 
For \cifar, $nPartitions = 64$, $ratio = 0.0046$; for \imagenet, $nPartitions = 64$, $ratio = 0.0074$. 
We compare DeepEverest against \textit{PreprocessAll} and \textit{ReprocessAll}. 
\cref{fig:performance} shows the results.

As the figures show, \textit{PreprocessAll} has the highest storage cost (37.8~GB for \cifar, and 1.35~TB for \imagenet) since it stores all activations for every input. However, scanning the precomputed activation values generally leads to the fastest query times. The query times of \textit{PreprocessAll} are slower for the early layer of \cifar because it has a large number of neurons, and thus it takes longer to load all the activations. \textit{ReprocessAll} has the lowest storage cost since it does not precompute or store anything ahead of time. Its query times are slow because of the DNN inference on the entire dataset at query time.

DeepEverest achieves the best of both worlds: low storage overhead and fast query times. For \cifar, compared with \textit{ReprocessAll} DeepEverest is \revision{$1.65{\times}$ to $31.1{\times}$} faster for \minor{\textit{FireMax}}, \revision{$1.22{\times}$ to $50.8{\times}$} faster for \minor{\textit{SimTop}}, and up to \revision{${31.5\times}$} faster for \minor{\textit{SimHigh}}. For \imagenet, compared with \textit{ReprocessAll}, DeepEverest is \revision{$2.67{\times}$ to $62.8{\times}$} faster for \minor{\textit{FireMax}}, \revision{$1.65{\times}$ to $63.5{\times}$} faster for \minor{\textit{SimTop}}, and \revision{$1.35{\times}$ to $63.1{\times}$} faster for \minor{\textit{SimHigh}}.

Compared to \textit{PreprocessAll} for both \cifar and \imagenet, DeepEverest achieves comparable and sometimes even faster query times for queries that target small and medium-size neuron groups despite using only $20\%$ of \textit{PreprocessAll}'s storage overhead. For queries that target large neuron groups, DeepEverest's query times are slower. 
We observe this phenomenon again in \cref{fig:budgetRobustness} (discussed in \cref{subsec:evalConfigSelect}). 
Due to the curse of dimensionality, there is little difference in the distances between different pairs of inputs. As a result, DeepEverest is not able to reduce the number of inputs run by the DNN at query time as it does for small and medium neuron groups. \cref{tab:nPartitionsNInputsRun} shows the number of inputs run by the DNN at query time to compute the activation values for \minor{\textit{SimHigh}} queries. We find that the number of inputs run by the DNN at query time for queries on larger neuron groups is higher than that of queries on smaller neuron groups.

\vspace{-0.95em}
\subsection{Multi-Query Workloads} \label{subsec:evalCumulativePerf}

In this section, we evaluate DeepEverest on multi-query workloads using incremental indexing described in \cref{subsec:incrementalIdx} that avoids start-up overhead. 
We construct various query workloads to represent possible DNN interpretation patterns and compare DeepEverest against other on-disk caching techniques. 
All workloads consist of $1{,}000$ the most general \minor{\textit{SimHigh}} queries that target neuron groups of medium size. 
The first query of each workload targets a \minor{\textit{RandHigh}} neuron group from a randomly selected layer.
Each later query has probabilities $p_{same}$ of querying the same layer as the previous query, $p_{prev}$ to query one of the previously queried layers (excluding the layer queried by the previous query), and $p_{new}$ to query a layer that has not been queried yet.
Workload 1 sets these to $p_{same}{=}0.5, p_{prev}{=}0.3, p_{new}{=}0.2$. 
Workload 2 sets these to $p_{same}{=}0.5, p_{prev}{=}0.4, p_{new}{=}0.1$.
Workloads 1 and 2 are intended to simulate the exploration process of users that are likely to initially target layers they are interested in, and gradually explore more layers. 
Additionally, we construct Workload 3 in which queries are independent of each other; each layer is targeted uniformly at random by each query.
This is not a realistic interpretation pattern but is meant to show the worst-case workload for DeepEverest.

We measure the cumulative total time, which includes \minor{the time for} both preprocessing and query execution, and cumulative storage for each method. DeepEverest is given a storage budget of $20\%$ of full materialization, and $nPartitions$ and $ratio$ are selected by our heuristic algorithm.
\textit{LRU Cache} and \textit{Priority Cache} have the same $20\%$ storage budget. %
The time to initially compute and store the data on disk is included with the $0$-th query for \textit{PreprocessAll} and \textit{Priority Cache}. 
The results for cumulative total time are shown in \cref{fig:CTTIncrementalIdx}. 
We report the storage results in the text.

DeepEverest consistently performs the best for Workloads 1 and 2 using less than $20\%$ of the storage of full materialization. 
We observe that after some number of queries, the cumulative total time of DeepEverest grows more slowly. 
For \cifar, it plateaus after around 300 queries for Workload 1 and around 550 queries for Workload 2. 
This indicates that DeepEverest has built and stored \minor{NPI} and \minor{MAI} for all layers in the DNN. 
All later queries are much faster because they benefit from these indexes and \minor{NTA}. 
For \imagenet, DeepEverest completes building and storing the indexes for all layers after around 780 queries for Workload 1 and never completes for Workload 2, as we observe that DeepEverest's storage consumption is only \revision{$11.3\%$} of full materialization after 1,000 queries. 
DeepEverest finishes building its indexes after fewer queries in Workload 1 than in Workload 2 since Workload 1 has a higher probability of querying new layers.

While DeepEverest has the fastest query times, its storage also grows more slowly than the baseline approaches (except for \textit{ReprocessAll} which does not have any storage overhead).
For both datasets and models, \textit{PreprocessAll} uses full storage after its preprocessing step. Similarly, \textit{Priority Cache} consumes its $20\%$ storage budget after preprocessing. \textit{LRU Cache} consumes its storage budget after around 50 to 200 queries.
As discussed above, DeepEverest finishes building the indexes for all layers and consumes its storage budget after around 300 to 500 queries on \cifar, and for \imagenet it fills its storage after 780 queries for Workload 1 and never does so for Workload 2.

For Workload 3, which is unlikely an interpretation pattern, DeepEverest is slightly worse than the best performing method for the first 200 to 300 queries on both datasets and models 
because during that time DeepEverest builds indexes for many new layers that have not been queried before. However, DeepEverest performs the best after 400 queries because more queries target previously seen layers and benefit from its indexes and NTA.

We further observe that users typically pause between queries. DeepEverest can use that time to compute and persist its indexes to disk, which would yield even better user-perceived query times.

\vspace{-0.3em}
\subsection{\revision{DeepEverest's Configuration Selection}} \label{subsec:evalConfigSelect}
\revision{We now study the effectiveness of NPI and MAI, and the impact of DeepEverest's configurable parameters, as well as how DeepEverest performs using the configuration selection heuristic described in \cref{subsubsec:configSelection} with different storage budgets.}

\nPartitionsEffect
\ratioEffect

\vspace{0.115em}
\noindent \textbf{Impact of Number of Partitions.} We first examine how the number of partitions, $nPartitions$, affects the query times. We measure the query times of DeepEverest on \minor{\textit{SimHigh}} queries after building \minor{NPI} with varying $nPartitions$. \minor{MAI} is disabled for this experiment. The results are shown in \cref{fig:nPartitionsEffect}. We also measure the number of inputs on which DeepEverest performs DNN inference during query processing. \cref{tab:nPartitionsNInputsRun} shows the results for \textit{CIFAR10-VGG16}. Similar trends are observed for \textit{ImageNet-ResNet50}.

The query time initially decreases as $nPartitions$ increases.
This is because when partitions are larger, inputs that do not contribute to the result end up being processed by DeepEverest.
Hence, as partitions get smaller, the number of inputs run by the DNN at query time decreases.
Then, after $nPartitions$ increases past a certain value (64 for \textit{CIFAR10-VGG16}; 128 for \textit{ImageNet-ResNet50}), the query time no longer decreases despite the number of inputs run by the DNN at query time continuing to decrease. 
Recall that \minor{NTA} runs inference on all inputs in a partition as it processes that partition. 
When $nPartitions$ is so large that the partition size is below the optimal $batchSize$ for DNN inference, the parallelism of the GPU is not fully utilized, which causes some queries to slow down. 
Therefore, a good value of $nPartitions$ creates partitions whose sizes are similar to the optimal $batchSize$.

\vspace{0.115em}
\noindent \textbf{Effectiveness of \minor{MAI}.} This experiment evaluates the effectiveness of \minor{MAI}. We measure the speedups of query times compared with \textit{ReprocessAll} when varying $ratio$, which determines the fraction of inputs with activation values materialized in \minor{MAI}. 
Recall that when \minor{MAI} is non-empty, it becomes the $0$-th partition.
For this experiment, we set $nPartitions = 16$, which is a setting that performs well (see \cref{fig:nPartitionsEffect}). 
As discussed in \cref{subsubsec:maxActIdx}, MAI is designed to accelerate \minor{\textit{FireMax}} and \minor{\textit{SimTop}} queries. 
We measure the speedups of such queries \minor{on neuron groups of different sizes}.

\cref{fig:ratioEffect} shows the results. 
Note that when $ratio {=} 0$, DeepEverest runs without \minor{MAI}. %
The speedups of query times are generally much higher when $ratio$ is any non-zero value. 
This is because \minor{MAI} enables DeepEverest to return the query results after processing a subset of the inputs \minor{from MAI (partition $0$)}, rather than processing the entire partition. 
We also observe that the speedups of query times plateau or drop as $ratio$ further increases. 
This is because loading \minor{MAI} from disk takes longer as $ratio$ increases. 
When a small index provides enough information for DeepEverest to find the top-$k$ results after processing only some inputs from MAI, increasing $ratio$ degrades the speedups; the additional inputs in MAI do not improve the query times, and loading a larger index takes longer.
The best value of $ratio$ in practice depends on the queries and the distributions of the activations of the queried neuron group. Empirically, we observe that a small value of $ratio$ (e.g., 0.05) is good for \minor{\textit{FireMax}} and \minor{\textit{SimTop}} queries on the two datasets and models.
\nPartitionsNInputsRun

\vspace{0.115em}
\noindent \textbf{Impact of Storage Budget.} 
In previous sections, we examined the performance of DeepEverest with a storage budget of $20\%$ of full materialization. Here we examine how well DeepEverest performs when the configuration selection algorithm has different storage budgets. 
We measure the speedups of query times compared with \textit{ReprocessAll} for \minor{\textit{SimTop}} and \minor{\textit{SimHigh}} queries that target medium and large neuron groups, as shown in \cref{fig:budgetRobustness}. 
We observe that empirically DeepEverest delivers high speedups across different storage budgets, which also suggests that our configuration selection algorithm is robust. 
With a larger storage budget, DeepEverest performs better. 
We also observe that the speedups of queries on medium neuron groups are generally greater than the speedups for queries on large neuron groups due to the curse of dimensionality.

\budgetRobustness

\vspace{-0.5em}
\subsection{Preprocessing Costs} \label{subsec:evalPreprocessing}

\revision{
This experiment evaluates the costs of preprocessing in an extreme case where the user queries all layers in the DNN. 
Note that with incremental indexing, the costs of index computation and persistence for each layer are incurred once \textit{only} if that layer is queried. 
In this experiment, DeepEverest is given a storage budget of $20\%$ of full materialization and preprocesses all layers for each dataset and model from the first layer to the last layer. 
}
Convolutional layers, activation layers, and batch normalization layers are considered separate layers. 
We measure the cumulative times for each component in preprocessing: DNN inference, data persistence (for \textit{PreprocessAll}, persisting the activations to disk; for DeepEverest, persisting \minor{NPI} and \minor{MAI} to disk), and index computation. 
We force-write the data to disk when measuring the time for data persistence. 
\cref{fig:preprocessingTime} shows the results.
DeepEverest has similar preprocessing times compared with \textit{PreprocessAll}.
The time for building \minor{NPI} and \minor{MAI} and persisting them to disk is similar to the time for \textit{PreprocessAll} to persist the activations to disk. 
Note that DNN inference takes longer for late layers than for early layers. 
We also observe that data persistence and index computation for early layers takes longer than for late layers, since the sizes of early layers are usually greater than that of late layers. 
Considering these results along with the results shown in \cref{subsec:evalEndToEnd}, DeepEverest achieves comparable and sometimes better query times than \textit{PreprocessAll}, with only $20\%$ of its storage overhead and similar preprocessing times.

\vspace{-0.4em}
\subsection{Effectiveness of \minor{IQA}} \label{subsec:evalInterQuery}

Finally, we evaluate the effectiveness of Inter-Query Acceleration \minor{(IQA)} for sequences of related queries. We randomly select five inputs and construct \revision{two} sequences of related queries for each input on various layers.
\revision{
Both sequences consist of $1{,}000$ \textit{SimHigh} queries. 
The first query of each sequence targets a \textit{RandHigh} neuron group containing $n_{size}$ neurons. Each later query randomly replaces $n_{replace}$ neurons in the neuron group of the previous query by $n_{replace}$ randomly selected \textit{RandHigh} neurons.
Sequence 1 sets $n_{size} {=} 5$ and $n_{replace} {=} 1$. Sequence 2 sets $n_{size} {=} 10$ and $n_{replace} {=} 2$.
}

We measure the speedups for query times of DeepEverest with IQA against DeepEverest without IQA for each query. 
\cref{fig:cifarInterQueryAcc} shows the median of the speedups for each query on \textit{CIFAR10-VGG16} given an \minor{in-memory} cache budget of \revision{1 GB} for \minor{IQA}, \minor{with $nPartitions{=}16$ and $ratio{=}0$ set for DeepEverest.} We observe that even with this small budget, \minor{IQA} consistently improves DeepEverest's query times across different layers. Not shown in the paper, we also experiment with \minor{different $nPartitions$ and $ratio$} \minor{and different cache budgets}. 
We find that \minor{IQA always} consistently speeds up related queries and larger budgets generally lead to larger speedups.
In \cref{fig:cifarInterQueryAcc}, the speedups for the first query are around $1{\times}$ since the in-memory cache is initially empty.
For later queries when the cache is populated, \revision{for Sequence 1, DeepEverest with IQA achieves speedups of $2.65{\times}$ to $8.73{\times}$ for the late layer, $3.97{\times}$ to $8.08{\times}$ for the mid layer, and $1.53{\times}$ to $3.38{\times}$ for the early layer.
For Sequence 2, DeepEverest with IQA achieves speedups of $4.00{\times}$ to $8.06{\times}$ for the late layer, $4.29{\times}$ to $7.86{\times}$ for the mid layer, and $1.48{\times}$ to $1.82{\times}$ for the early layer.} 
The speedups for the early layer are smaller because this layer is larger, and hence the in-memory cache can hold fewer inputs' activations of the full layer. 
We observe larger speedups for the early layer with larger cache budgets.

\preprocessingTime
\cifarInterQueryAcc

\section{Discussion} \label{sec:discussion}
This section discusses some possible optimizations and extensions for DeepEverest.

\noindent \textbf{Incrementally Returning Query Results.} NTA runs until it has found $k$ inputs whose distances to the sample $s$ are at most the threshold value, $t$.
However, NTA may be certain that some inputs are part of the top-$k$ set before it has found the complete set. For queries where $k > 1$, after each round of the algorithm, DeepEverest returns inputs in $Y \subseteq U$, where for all $y \in Y$, $dist(s, y, G) \le t$, and continues running to find the rest of the $k - |Y|$ results.
DeepEverest's optimizations enable it to incrementally return the top-$k$ results quickly, and therefore reduces the time required to return the first part of the answer to the user.

\noindent \textbf{Approximation.} Modifying DeepEverest to give approximate results is straightforward. Following the definition of the $\theta$-approximation in Fagin's paper \cite{fagin2003optimal}, a $\theta$-approximation (let $0 < \theta < 1$ be given) to the top-$k$ answers is a collection of $k$ inputs, $U$, (and their distances to the sample input) such that for each $y \in U$ and each $z \in D \setminus U$, $\theta * dist(s, y, G) \le dist(s, z, G)$. Let $t$ be the threshold value from \cref{eq:threshold}. DeepEverest can find a $\theta$-approximation to the top-$k$ answers by modifying the termination condition in \cref{eq:termination} to be,
\useshortskip
\begin{equation} \label{eq:termination-approx}
    \max_{x \in top} \left\{ dist(s, x, G) \right\} \leq t / \theta
\end{equation}

\noindent \textbf{Early Stopping.} DeepEverest can be further modified into an interactive process in which it can show the user the current top-$k$ results with a guarantee about the degree of approximation to the correct top-$k$ results. Based on this guarantee, the user can decide whether they would like to stop the process at any time. Let $b$ be the largest distance to the sample input from the current top-$k$ results, let $t$ be the current threshold value, and let $\theta = t / b$. If the algorithm is stopped early, we have $0 < \theta < 1$ because $b > t$. Therefore, the current top-$k$ results is then a $\theta$-approximation to the correct top-$k$ answers. Thus, the user can be shown the current top-$k$ results and the number $\theta$, with a guarantee that they are being shown a $\theta$-approximation.

\vspace{-0.2em}
\section{Conclusion}
We presented DeepEverest, a system that accelerates top-$k$ queries for DNN interpretation. DeepEverest, with various optimizations, reduces the number of activations computed at query time with low storage overhead, while guaranteeing correct query results. With less than $20\%$ of the storage of full materialization, DeepEverest accelerates individual queries by up to \revision{$63.5{\times}$} and consistently outperforms other methods over various multi-query workloads.

\section*{Acknowledgments}
This work was supported in part by NSF grants OAC-1739419 and OAC-1934292.

\balance

\end{sloppypar}

\bibliographystyle{ACM-Reference-Format}
\bibliography{header,others,self}
\balance

\end{document}